\documentclass[11pt,letterpaper]{article}

\usepackage{qc}
\usepackage[letterpaper,margin=1in]{geometry}
\usepackage{authblk}
\usepackage{csquotes} 
\usepackage[backend=biber,style=alphabetic,maxbibnames=99]{biblatex}
\usepackage[capitalise,nameinlink,noabbrev]{cleveref}

\newcommand{\dist}{\operatorname{dist}}

\definecolor{RevisionBlue}{RGB}{0,0,0}
\newcommand{\rev}[1]{\textcolor{RevisionBlue}{#1}}
\newenvironment{revision}{\begingroup\color{RevisionBlue}}{\endgroup}
\AtBeginBibliography{\tolerance=1500\setlength{\emergencystretch}{2em}}

\title{Quantum Copy Complexity of Testing Monotonicity \\ and Other Properties}
\author{Preksha Naik\thanks{\normalsize\texttt{pnaik@caltech.edu}}}
\affil{California Institute of Technology}
\date{\today}

\begin{document}
\maketitle

\begingroup
\let\small\normalsize
\begin{abstract}
We study the sample complexity of testing monotonicity of Boolean functions when the tester is given only copies of the quantum function state $\ket{\psi_f}=\frac{1}{\sqrt{2^n}}\sum_{x\in\{0,1\}^n}\ket{x,f(x)}$. We give a tester with sample complexity $O\!\left(\min\left\{\frac{n}{\varepsilon^3}, \frac{n^{3/2}}{\varepsilon^2}\right\}\right)$, improving on the previous $\widetilde O(n^2/\varepsilon^2)$ upper bound of~\cite{caro2025testingclassicalpropertiesquantum}. We also prove the first nontrivial lower bounds for monotonicity testing from quantum function states, matching our upper bounds over a broad range of $\varepsilon$.

Our upper bounds use two approaches. The first approach refines the Fourier-based tester of~\cite{caro2025testingclassicalpropertiesquantum} by handling high-influence functions separately, adapting an idea from~\cite{DBLP:journals/siamcomp/KhotMS18}. The second approach uses the classical path tester of~\cite{DBLP:journals/siamcomp/KhotMS18} by constructing a quantum observable whose expectation equals the path tester's violation probability.

To prove the lower bounds, we place independent functions in mutually incomparable regions of the Boolean hypercube, building on the classical constructions of~\cite{belovs2016polynomial,chen2017talagrand}. We introduce a new argument for the quantum setting that replaces a fixed number of copies with a Poisson-distributed number of copies, allowing the contributions from different regions to be analyzed independently.

This same framework also allows us to prove an $\Omega(n^2)$ lower bound for testing union-closedness and $2$-monotonicity from quantum function states.
\end{abstract}
\endgroup

\clearpage
\tableofcontents
\clearpage

\section{Introduction}\label{sec:introduction}

In property testing, we study whether a small amount of information suffices to decide if an unknown object satisfies a property or is far from satisfying the property. More formally, given a collection $\mathcal P$ of Boolean functions and a parameter $\varepsilon>0$, the goal is to distinguish between 
\[
    f\in\mathcal P
    \qquad\text{and}\qquad
    \dist(f,\mathcal P)\geq\varepsilon,
\]
promised that one of these is the case. Here,
\[
    \dist(f,\mathcal P)
    :=\min_{g\in\mathcal P}\dist(f,g),
    \qquad
    \dist(f,g)
    :=\Pr_{x\sim\mathrm{Unif}(\{0,1\}^n)}
       [f(x)\neq g(x)].
\]

Monotonicity testing on the Boolean hypercube has served as a canonical example for property testing algorithms, both in the classical and quantum setting. For example, see~\cite{Goldreich2000,fischer2002posets,DBLP:journals/siamcomp/KhotMS18,Belovs2015,black_nearly_2023,caro2025testingclassicalpropertiesquantum}. A Boolean function $f:\{0,1\}^{n}\to\{0,1\}$ is monotone if
\[
    f(x) \leq  f(y)
    \qquad\text{whenever }x \preceq y,
\]
where $x \preceq y$ denotes the coordinate-wise order: $x_i\leq y_i$ for every $i\in[n]$. The goal is to determine whether a function that is far from monotone can be detected given only limited access to the function. 

The answer depends on how the tester is able to access our function $f$. In the classical query model, a tester is able to choose any input and observe its function value while in the classical sample-based model, a tester receives random labeled examples $(x,f(x))$ (often drawn uniformly). In the quantum query model, the tester can apply the oracle $|x,b\rangle\mapsto|x,b\oplus f(x)\rangle$ to any state coherently. In this paper, we study the quantum sample-based setting, where the tester receives copies of the uniform quantum state
\begin{equation}
\label{eq:uniform_function_state}
    \ket{\psi_f}=2^{-n/2}\sum_{x\in\{0,1\}^n}\ket{x,f(x)}.
\end{equation}
The tester may perform arbitrary measurements on multiple copies of this state. Our goal is to minimize the number of copies needed to test any property.

Quantum states of the form $\sum_x \sqrt{D(x)}\ket{x,f(x)}$ were first studied in the context of PAC learning in~\cite{bshouty1995learning}. The state in \Cref{eq:uniform_function_state} corresponds to taking $D$ to be uniform. Quantum examples and related models  have since been studied in both learning and property testing  ~\cite{atici2007quantum,arunachalam2018samplecomplexity,cross2015quantumlearning,arunachalam2023optimal}. Most closely related to our work is that of ~\cite{caro2025testingclassicalpropertiesquantum} who give a monotonicity tester that uses $\widetilde O(n^2/\varepsilon^2)$ copies. The authors also show that for monotonicity, the hard distributions used for classical lower bounds can be distinguished using only a few quantum function states, yielding a trivial lower bound of $\Omega(1/\varepsilon)$ copies. In this work, we close this gap by giving matching upper and lower bounds for monotonicity testing in the quantum sample model. \Cref{tab:access-models} places our bounds alongside the best known results in other access models.

\begin{table}[htbp]
\centering
\renewcommand{\arraystretch}{1.45}
\begin{tabular}{@{}
    >{\raggedright\arraybackslash}p{0.22\textwidth}
    >{\raggedright\arraybackslash}p{0.34\textwidth}
    >{\raggedright\arraybackslash}p{0.34\textwidth}
@{}}
\hline
Access model & Upper bounds & Lower bounds \\
\hline

Classical samples
& $2^{\widetilde O(\sqrt n/\varepsilon)}$
\par\cite{bshouty1996fourier,black_nearly_2023}
& $2^{\Omega(\sqrt n/\varepsilon)}$
\par\cite{black_nearly_2023}
\\[5pt]

Classical queries
& $\widetilde O\!\left(\frac{\sqrt n}{\varepsilon^2}\right)$
\par\cite{DBLP:journals/siamcomp/KhotMS18}
& $\Omega(n^{1/2-\eta})$
\par for any fixed $\eta>0$
\par\cite{chen2026adaptive}
\\[5pt]

Quantum queries
& $\widetilde O\!\left(\frac{n^{1/4}}{\sqrt\varepsilon}\right)$
\par\cite{Belovs2015}
& ---
\\[5pt]

Quantum examples
& $O\!\left(\min\left\{
\frac{n}{\varepsilon^3},
\frac{n^{3/2}}{\varepsilon^2}
\right\}\right)$
\par (this work)
& $\Omega\!\left(\min\left\{
\frac{n}{\varepsilon^3},
\frac{n^{3/2}}{\varepsilon^2}
\right\}\right)$
\par (this work)
\\

\hline
\end{tabular}
\caption{Bounds for monotonicity testing across different access models. Our lower bound holds for $2^{-\sqrt n/200}\leq\varepsilon\leq1/128$.}
\label{tab:access-models}
\end{table}

\subsection{Our results}\label{sec:results}
The following statements summarize the results proved in this paper. 

\begin{theorem}[Upper bound]
\label{thm:intro-upper}
For every $n\geq1$, $0<\varepsilon\leq1/2$, and $0<\delta<1/2$, there exists a quantum algorithm that uses at most
\[
    O\!\left(
        \min\left\{
            \frac{n^{3/2}}{\varepsilon^2},
            \frac{n}{\varepsilon^3}
        \right\}
        \log\frac{1}{\delta}
    \right)
\]
copies of the state $\ket{\psi_f}$ to decide, with success probability at least $1-\delta$, whether $f$ is monotone or $\varepsilon$-far from monotone.
\end{theorem}

\begin{theorem}[Lower bound]
\label{thm:intro-lower}
For $2^{-\sqrt n/200}\leq\varepsilon\leq1/128$,
any quantum algorithm requires
\[
    \Omega\!\left(\min\left\{
        \frac{n}{\varepsilon^3},
        \frac{n^{3/2}}{\varepsilon^2}
    \right\}\right)
\]
copies of $\ket{\psi_f}$ to decide, with success probability at least $2/3$, whether $f$ is monotone or $\varepsilon$-far from monotone.
\end{theorem}

The lower-bound construction can also be applied to at least two other Boolean properties. In particular, a function is union-closed if $f(x)=f(y)=1$ implies $f(x\vee y)=1$, where $x\vee y$ denotes coordinate-wise OR. Furthermore, a function is $2$-monotone if there do not exist $x\prec y\prec z$ such that $f(x)=1$, $f(y)=0$, and $f(z)=1$ where $x\prec y$ refers to coordinate-wise order. We show that the following lower bounds apply.  

\begin{theorem}[Additional lower bounds]
\label{thm:intro-other}
For every $0<\varepsilon\leq1/256$, any quantum algorithm requires $\Omega(n^2)$ copies of the state $\ket{\psi_f}$ to decide, with success probability at least $2/3$,
whether $f$ is union-closed or $\varepsilon$-far from union-closed.
The same holds for testing $2$-monotonicity. 
\end{theorem}

\subsection{Overview of our proofs}\label{sec:techniques}

\paragraph{Upper bounds.} We prove our upper bound using two separate testers. 

Our first tester improves on the quantum version of the classical edge tester. The classical edge tester of~\cite{Goldreich2000} samples a uniformly random edge $(x,y)$ of the Boolean hypercube and checks whether it is violated. A monotone function has no violated edges, while a function that is $\varepsilon$-far from monotone has a nontrivial probability of violated edges.  ~\cite{caro2025testingclassicalpropertiesquantum} estimate this probability using copies of the quantum function state. Specifically, the probability that a random edge of the Boolean hypercube violates monotonicity can be written using the total influence of the function as well as the first-level Fourier coefficients. The influence can be estimated using Fourier sampling while the first-level coefficients can be estimated from measurements in the computational basis.

Our improvement is to first filter functions that have large influence. This idea appears in the classical query-based monotonicity tester of~\cite{DBLP:journals/siamcomp/KhotMS18}. Since every monotone function has influence at most $\sqrt n$, we can first do a coarse estimate to reject functions that have influence much larger than this. In the low-influence regime, the estimator for the influence has lower variance which improves the quantum copy complexity.  

Our second tester is based on the classical path tester of 
~\cite{DBLP:journals/siamcomp/KhotMS18}. While the edge tester only looks at neighboring endpoints of the hypercube, the path tester looks at comparable points farther away. The classical path tester samples a starting point and moves either upward or downward along a random set of coordinates and then checks whether the two endpoints form a violation. 

Since we cannot query function values, we construct quantum observables whose expectations reproduce the violation probability of the classical path tester for a specific path length. We estimate these quantities at several path lengths. The structural results of~\cite{DBLP:journals/siamcomp/KhotMS18}, together with the stronger directed isoperimetric inequality of~\cite{pallavoor2022distance}, guarantee that some scale has enough violations whenever the function is far from monotone. The stronger inequality is primarily used to eliminate a logarithmic factor in the final bound. Furthermore, as in the Fourier tester, we do a coarse estimate to eliminate functions with high influence, allowing the relevant estimators to have small variance. This gives a tester that uses $O(n/\varepsilon^3)$ copies.

These two testers together establish the upper bound in \Cref{thm:intro-upper}.

\paragraph{Lower bounds.}
For the lower bound, we construct two distributions over Boolean functions: one supported on monotone functions and one supported on functions that are $\varepsilon$-far from monotone with high probability. We then show that the corresponding averaged $t$-copy states are difficult to distinguish, even using arbitrary measurements on all copies by bounding the trace distance.

The construction is based on ideas from classical monotonicity lower bounds~\cite{belovs2016polynomial,chen2017talagrand}. We select many small, pairwise incomparable regions of the hypercube and place an independent local function in each region. Outside of these regions, we assign function values that preserve monotonicity whenever the local functions are monotone. Because no point in one region is comparable to a point in another, the monotonicity behavior of different regions do not interfere. The global function is built out of many independent local pieces.

Next, within each region, we choose a smaller band where the local function can vary. Restricting the construction to this band lets us control how much each function contributes to the global distance from montonocity. Outside of the band, all of the local functions are identical. Inside the band, we use dictators and constant functions for the monotone case. In the far from monotone case, we use reversed dictators to create violations. The width of the band determines how large the local contribution to the distance can be, while the probability of choosing a reversed dictator determines how often such violations occur. By tuning these two parameters, we make the global function $\varepsilon$-far from monotone with high probability.

The main difficulty now is to show that the two distributions remain hard to distinguish when the tester receives copies of the quantum function states. Let $\mathcal{D}_{\mathrm{yes}}$ and $\mathcal{D}_{\mathrm{no}}$ denote the two distributions. We want to bound
\[
\operatorname{TD}\!\left(
\mathbb{E}_{f\sim\mathcal{D}_{\mathrm{yes}}}
\left[(\ket{\psi_f}\!\bra{\psi_f})^{\otimes t}\right],
\mathbb{E}_{f\sim\mathcal{D}_{\mathrm{no}}}
\left[(\ket{\psi_f}\!\bra{\psi_f})^{\otimes t}\right]
\right).
\]

Classically, the independence between the different regions can be used directly. In the quantum setting, however, the global $t$-copy state does not factor across the regions. To overcome this, we introduce an auxiliary state that is a coherent superposition over different copy numbers, with weights corresponding to a Poisson distribution. In this representation, overlaps between global states factor across the regions. This allows us to analyze each region separately and then combine the resulting bounds. Finally, we transfer the bound back to the original setting with a fixed number of copies.

We consider two ranges of $\varepsilon$. When $n^{-1/2}\leq\varepsilon\leq1/128$, we choose the band inside each region to have mass $\Theta(\varepsilon)$, which gives an $\Omega(n/\varepsilon^3)$ lower bound. When $2^{-\sqrt n/200}\leq\varepsilon<n^{-1/2}$, we instead keep the band at mass $\Theta(n^{-1/2})$ and reduce the probability of choosing a reversed dictator. This gives an $\Omega(n^{3/2}/\varepsilon^2)$ lower bound. Together, the two regimes match our upper bounds over the stated range of $\varepsilon$.

\paragraph{Other properties.}
The same lower bound framework can also be applied to union-closedness and $2$-monotonicity. In particular, we can keep the same incomparable regions and reuse the quantum indistinguishability argument. However, we replace dictators and the reversed dictators by XOR and XNOR functions on pairs of coordinates. This gives the $\Omega(n^2)$ copy lower bounds for both properties.

\section{Preliminaries}\label{sec:preliminaries}

\paragraph{Notation and distance.}

We write $[n]:=\{1,\ldots,n\}$ and $N:=2^n$. Furthermore, unless stated otherwise, all Boolean functions take values in $\{0,1\}$ and probabilities and expectations over $\{0,1\}^n$ are computed over the uniform distribution. For $x\in\{0,1\}^n$, we write $|x|:=\sum_{i=1}^n x_i$ for its Hamming weight. 

We write $x\preceq y$ if $x_i\leq y_i$ for every
$i\in[n]$, and denote by $\mathrm{MON}_n$ the collection
of Boolean functions satisfying $f(x)\leq f(y)$
whenever $x\preceq y$. For any comparable pair, we call $x$ the lower endpoint and $y$ the upper endpoint. A pair is a violation if $f(x) = 1$ and $f(y) = 0$. Furthermore, it is a violated edge if the endpoints specifically differ in exactly one coordinate. 

\paragraph{Quantum examples and copy complexity.}
For a Boolean function $f$, let $|\psi_f\rangle$ be the function state in \Cref{eq:uniform_function_state}. We use the terms \emph{quantum function state},
\emph{quantum example}, and \emph{quantum copy} interchangeably. A tester receiving $t$ copies may perform an arbitrary joint measurement on $|\psi_f\rangle^{\otimes t}$.

\paragraph{Quantum distance conventions.}
For density operators $\rho$ and $\sigma$, we use
trace distance with the normalization
\[
    \operatorname{TD}(\rho,\sigma)
    :=\frac12\|\rho-\sigma\|_1,
\]
and use the unsquared fidelity
\[
    F(\rho,\sigma)
    :=\|\sqrt{\rho}\sqrt{\sigma}\|_1.
\]
Furthermore, $\|A\|_1:=\operatorname{Tr}\sqrt{A^\dagger A}$ is the trace norm. For density operators,  we have that $\operatorname{TD}(\rho,\sigma)\leq \sqrt{1-F(\rho,\sigma)^2}$~\cite{fuchs1999cryptographic}. Furthermore, the optimal success probability for distinguishing two equally likely states is $(1+\operatorname{TD}(\rho,\sigma))/2$~\cite[Theorem~3.4]{watrous2018theory}.
For a distribution $\mathcal D$ over functions, the averaged $t$-copy state is given by $\mathbb{E}_{f\sim\mathcal D}\left[(|\psi_f\rangle\langle\psi_f|)^{\otimes t}\right]$.

\section{Upper bounds on monotonicity testing}\label{sec:upper-bounds}

We give two algorithms for testing monotonicity from quantum function states, each with a different tradeoff between the dimension $n$ and the distance parameter $\varepsilon$. 

\subsection{Fourier-based tester}
\label{sec:fourier-upper}

Our first algorithm refines the Fourier-based monotonicity tester introduced in~\cite[Section~2.2]{caro2025testingclassicalpropertiesquantum}. We follow their notation and recall the relevant definitions and identities before presenting the improvement. 

Let $g:\{-1,1\}^n\to\{-1,1\}$ be the sign representation of $f$, obtained by relabeling $0$ as $1$ and $1$ as $-1$ in both the input and output. For $i\in[n]$, let $z^{\oplus i}$ denote $z$ with its $i$-th coordinate flipped, and define
\[
    \operatorname{Inf}_i[g]
    :=\Pr_z[g(z)\neq g(z^{\oplus i})],
    \qquad
    \mathbf I[g]
    :=\sum_{i=1}^n\operatorname{Inf}_i[g].
\]
Furthermore, we use 
\[
    \widehat g(S)
    :=\mathbb E_z\left[g(z)\prod_{i\in S}z_i\right]
\]
for the Fourier coefficients of $g$.

Let $p$ denote the probability that a uniformly random edge of the Boolean hypercube violates monotonicity. In particular, 
\[
    p:=\Pr_{x\sim\{0,1\}^n,\,i\sim[n]}
    \left[
        (x_i=0\wedge f(x)=1\wedge f(x+e_i)=0)
        \vee
        (x_i=1\wedge f(x)=0\wedge f(x^{\oplus i})=1)
    \right],
\]
where $x$ and $i$ are chosen independently and uniformly. As shown in~\cite[Section~2.2]{caro2025testingclassicalpropertiesquantum}, we have that
\begin{equation}\label{eq:fourier-violation-probability}
    p
    =
    \frac{1}{2n}
    \left(
        \mathbf I[g]
        -
        \sum_{i=1}^n\widehat g(\{i\})
    \right).
\end{equation}
Furthermore, the classical edge-tester bound (see ~\cite{Goldreich2000}) implies that
\begin{equation}\label{eq:fourier-testing-gap}
    f\in\mathrm{MON}_n
    \Longrightarrow p=0,
    \qquad
    \dist(f,\mathrm{MON}_n)\geq\varepsilon
    \Longrightarrow
    p\geq\frac{\varepsilon}{n}.
\end{equation} 
The algorithm in~\cite{caro2025testingclassicalpropertiesquantum}
gives unbiased estimators for both terms in~\eqref{eq:fourier-violation-probability}. Our improvement will use the additional fact that every monotone Boolean function satisfies
\begin{equation}\label{eq:monotone-influence-bound}
    \mathbf I[g]\leq\sqrt n;
\end{equation}
see~\cite[Section~2.3]{odonnell2021analysisbooleanfunctions}. We combine this with improved variance bounds for the estimators.

\begin{proposition}
\label{prop:fourier-upper}
There exists a quantum algorithm that, given $O\left(\frac{n^{3/2}\log(1/\delta)}{\varepsilon^2}\right)$ copies of
$\ket{\psi_f}=\frac{1}{\sqrt{2^n}} \sum_{x\in\{0,1\}^n} \ket{x,f(x)}$
distinguishes monotone functions from functions that are $\varepsilon$-far from monotone with probability at least $1-\delta$ for $0<\delta<1/2$.
\end{proposition}

\begin{proof}
We first construct a tester with error probability at most $1/3$, and we reduce the error to $\delta$ by repeating the tester. The algorithm proceeds in two stages. We first perform a coarse estimate  $\widehat{\mathbf I}_0$ of the influence $\mathbf I[g]$ and reject if
$\widehat{\mathbf I}_0>\frac32\sqrt n$. If this step does not reject, we estimate $p$ using
the expression in~\eqref{eq:fourier-violation-probability}
and the estimators
from~\cite[Section~2.2]{caro2025testingclassicalpropertiesquantum}. Filtering on the high-influence case is also part of the classical argument in~\cite[Section~7.3]{DBLP:journals/siamcomp/KhotMS18}; here, this estimate is obtained from Fourier samples. 

We begin by bounding the variance of these estimators. Specifically, to estimate the total influence $\mathbf I[g]$,~\cite{caro2025testingclassicalpropertiesquantum} use Fourier-sampling. Apply $H^{\otimes(n+1)}$ to a quantum function state and measure in the computational basis. The last qubit is $1$ with probability $1/2$, and conditional on this
outcome, the first $n$ qubits give a set $S\subseteq[n]$ with probability $\widehat g(S)^2$. Define $Y=2|S|$ when the last qubit is $1$, and $Y=0$ otherwise. Then, $\mathbb E[Y]=\mathbf I[g]$. Furthermore, we have that
\begin{equation}\label{eq:fourier-influence-variance}
    \operatorname{Var}(Y)
    \leq\mathbb E[Y^2]
    =2\sum_{S\subseteq[n]}|S|^2\widehat g(S)^2
    \leq2n\mathbf I[g],
\end{equation}
where we have used that $|S| \leq n$ in the last inequality. 

Next, to estimate $\sum_{i=1}^n\widehat g(\{i\})$, ~\cite{caro2025testingclassicalpropertiesquantum} measure the quantum function state in the computational basis. This gives a uniformly random labeled example $(z, g(z))$ (in the sign representation). The estimator for $\widehat g(\{i\})$ is $g(z)z_i$ so we can define
\[
    Z:=g(z)\sum_{i=1}^n z_i.
\]
Then, we have $\mathbb E[Z]=\sum_{i=1}^n\widehat g(\{i\})$. Furthermore, we get
\begin{equation}\label{eq:fourier-first-level-variance}
    \begin{aligned}
        \operatorname{Var}(Z)
        &\leq \mathbb E[Z^2]
        =\mathbb E_z\left[\left(\sum_{i=1}^n z_i\right)^2\right]\\
        &=\sum_{i=1}^n\mathbb E_z[z_i^2]
          +\sum_{i\neq j}\mathbb E_z[z_i z_j]
        =n.
    \end{aligned}
\end{equation}
where we used that $g(z)^2=1$ for all $z$ and $z_i$ and $z_j$ are independent for all $i \neq j$. 

We now analyze the first stage of the algorithm. Let $C > 0$ be a constant to be chosen later, and let $m_0=\lceil C\sqrt n\rceil$. Using $m_0$ independent samples of $Y$, define 
\[
    \widehat{\mathbf I}_0
    :=\frac{1}{m_0}\sum_{r=1}^{m_0}Y_r.
\]
Then, we have that
\[
    \mathbb E[\widehat{\mathbf I}_0]=\mathbf I[g],
    \qquad
    \operatorname{Var}(\widehat{\mathbf I}_0)
    \leq\frac{2n\mathbf I[g]}{m_0}.
\]
The tester rejects if
$\widehat{\mathbf I}_0>\frac32\sqrt n$. If $f$ is monotone, then $\mathbf I[g]\leq\sqrt n$ by~\eqref{eq:monotone-influence-bound}. By Chebyshev's inequality, 
\begin{equation}\label{eq:fourier-filter-completeness}
    \Pr\left[
        \widehat{\mathbf I}_0>\frac32\sqrt n
    \right]
    \leq
    \frac{2n\mathbf I[g]}
         {m_0(\sqrt n/2)^2}
    \leq\frac8C.
\end{equation}
Therefore, a function that is monotone succeeds with probability at least $1-8/C$. On the other hand, if $\mathbf I[g]\geq2\sqrt n$, then $\mathbf I[g]-\frac32\sqrt n\geq\mathbf I[g]/4$. Applying Chebyshev's inequality once more, we have 
\begin{equation}\label{eq:fourier-filter-soundness}
    \begin{aligned}
        \Pr\left[
            \widehat{\mathbf I}_0\leq\frac32\sqrt n
        \right]
        &\leq
        \frac{2n\mathbf I[g]}
             {m_0(\mathbf I[g]-\frac32\sqrt n)^2}\\
        &\leq
        \frac{32n}{m_0\mathbf I[g]}
        \leq\frac{16}{C}.
    \end{aligned}
\end{equation}
Hence, functions with influence at least $2\sqrt n$ are rejected with probability at least $1-16/C$ in the first stage of the algorithm. We now analyze the second stage. 

If the first stage has not rejected, we use $m_I=\left\lceil\frac{Cn^{3/2}}{\varepsilon^2}\right\rceil$ copies to obtain samples of $Y$, whose average we can denote by $\widehat{\mathbf I}$. Furthermore, we use
$m_G=\left\lceil\frac{Cn}{\varepsilon^2}\right\rceil$
fresh copies to obtain uniformly random labeled examples. Averaging the corresponding values of $Z$ gives the estimate $\sum_{i=1}^n\widetilde g_i$ of $\mathbb E[Z]$,
where
\[
    \widetilde g_i
    :=\frac{1}{m_G}\sum_{r=1}^{m_G}
        g(z^{(r)})z_i^{(r)},
    \qquad i\in[n].
\]
We can then compute
\[
    \widehat p
    :=\frac{1}{2n}\left(
        \widehat{\mathbf I}
        -\sum_{i=1}^n\widetilde g_i
    \right),
\]
and reject if $\widehat p>\varepsilon/(2n)$. 

To analyze the remaining case, suppose that $\mathbf I[g]<2\sqrt n$. We have by \Cref{eq:fourier-influence-variance}, 
\[
    \operatorname{Var}(\widehat{\mathbf I})
    \leq\frac{2n\mathbf I[g]}{m_I}
    \leq\frac{4n^{3/2}}{m_I}
    \leq\frac{4\varepsilon^2}{C}.
\]
Furthermore, from \Cref{eq:fourier-first-level-variance}, we have 
\[
    \operatorname{Var}\left(
        \sum_{i=1}^n\widetilde g_i
    \right)
    \leq\frac{n}{m_G}
    \leq\frac{\varepsilon^2}{C}.
\]
Therefore,
\[
    \mathbb E[\widehat p]=p,
    \qquad
    \operatorname{Var}(\widehat p)
    \leq\frac{1}{4n^2}\left(
        \frac{4\varepsilon^2}{C}
        +\frac{\varepsilon^2}{C}
    \right)
    =\frac{5\varepsilon^2}{4Cn^2}.
\]
Applying Chebyshev's inequality, we get
\begin{equation}\label{eq:fourier-estimation-error}
    \Pr\left[
        |\widehat p-p|>\frac{\varepsilon}{4n}
    \right]
    \leq
    \frac{5\varepsilon^2/(4Cn^2)}
         {\varepsilon^2/(16n^2)}
    =\frac{20}{C}.
\end{equation}
If $f$ is monotone, then $p=0$. Whenever
$|\widehat p-p|\leq\varepsilon/(4n)$, we have
\[
    \widehat p\leq\frac{\varepsilon}{4n}
    <\frac{\varepsilon}{2n},
\]
and we accept. If $f$ is $\varepsilon$-far from monotone, then $p\geq\varepsilon/n$
by \Cref{eq:fourier-testing-gap}. Therefore, \[
    \widehat p
    \geq p-\frac{\varepsilon}{4n}
    \geq\frac{3\varepsilon}{4n}
    >\frac{\varepsilon}{2n},
\]
and the second stage rejects.

Combining the two stages, the probability of incorrectly
rejecting a monotone function is, by
\Cref{eq:fourier-filter-completeness}
and \Cref{eq:fourier-estimation-error}, at most
\[
    \frac8C+\frac{20}{C}=\frac{28}{C}.
\]
Next, if a function is $\varepsilon$-far from monotone and it satisfies $\mathbf I[g]\geq2\sqrt n$, the probability that it gets accepted is at most $16/C$ by \Cref{eq:fourier-filter-soundness}. For a function that is $\varepsilon$-far from monotone and has
$\mathbf I[g]<2\sqrt n$, the probability that it gets accepted is at most $20/C$ by \Cref{eq:fourier-estimation-error}. Therefore, choosing $C=84$ makes the error probability at most $1/3$ in every promised case.

To achieve error probability at most $1/3$, the algorithm uses at most
\[
    m_0+m_I+m_G
    =O\left(
        \sqrt n+\frac{n^{3/2}}{\varepsilon^2}
        +\frac{n}{\varepsilon^2}
    \right)
    =O\left(\frac{n^{3/2}}{\varepsilon^2}\right)
\]
copies. Finally, by Hoeffding's inequality, repeating the algorithm independently $O(\log(1/\delta))$
times and taking the majority outcome reduces the error probability to at most $\delta$. The total copy complexity is therefore
\[
    O\left(
        \frac{n^{3/2}\log(1/\delta)}{\varepsilon^2}
    \right).
\]

\end{proof}

\subsection{Path-based observable}
\label{sec:path-upper}

We first describe the tester and then prove correctness and bound the copy complexity in subsequent sections. For a constant $n_0$, all $n\geq n_0$, and $n^{-1/2}\leq\varepsilon\leq1/2$, let
\[
\mathcal D_n=\{1,2,4,\ldots,2^{\lfloor\log_2 n\rfloor}\},
\]
and for each $d\in\mathcal D_n$ define
\[
t_d=\max\{1,\lfloor c_0\varepsilon\sqrt n\,d^{-5/8}\rfloor\},
\qquad
a_d=\frac{c_1\varepsilon t_d d^{3/8}}{n}.
\]
The constants $c_0,c_1$ are fixed in \Cref{lem:path-soundness}. Furthermore, consider the observables $Q_d$ and $H_d\otimes Z$, defined in \Cref{sec:ordinary-path-observable} which depend only on $n,\varepsilon$, and $d$. If $U_d$ and $V_d$ denote their measurement outcomes, then we show that
\[
\mathbb E\!\left[\frac12U_d-\frac14V_d\right]=p_{t_d}(f),
\]
where $p_{t_d}(f)$ is the violation probability of the corresponding path experiment. The tester proceeds as follows.
\begin{enumerate}
\item Use
\[
m_0=\lceil96\sqrt n\rceil
\]
copies to obtain the coarse influence estimate from \Cref{sec:fourier-upper}. Reject if
\[
\widehat{\mathbf I}_0>\frac32\sqrt n.
\]

\item For each $d\in\mathcal D_n$, let
\[
m_d=\left\lceil
\frac{64C_{\mathrm{est}}}{c_1^2}
\frac{n^{3/2}}{\varepsilon^2t_dd^{3/4}}
\right\rceil,
\qquad
r_d=\min\{r\in\mathbb N:r\text{ is odd and }r\geq8\log(12d)\},
\]
where $C_{\mathrm{est}}$ is the variance constant from
\eqref{eq:ordinary-path-estimator-variance}.

For each of the $r_d$ repetitions, measure $Q_d$ on $m_d$ fresh copies and $H_d\otimes Z$ on another $m_d$ fresh copies. Let $\overline U_d$ and $\overline V_d$ denote the corresponding empirical means, and set
\[
\widehat p_d=\frac12\overline U_d-\frac14\overline V_d.
\]
Let $\widetilde p_d$ be the median of the resulting $r_d$ estimates.

\item Reject if
\[
\widetilde p_d\geq \frac{a_d}{2}
\]
for some $d\in\mathcal D_n$. Otherwise, accept.
\end{enumerate}

All of the measurements use independent copies, including the preliminary influence estimate. We will show that the tester above has constant error. To reduce the error to $\delta$, we repeat the entire procedure independently and take a majority vote.

\subsubsection{Structural bound and path analysis}

We begin by stating the isoperimetric
inequality of~\cite[Theorem~2.9]{pallavoor2022distance},
which we will use to refine the path analysis
of~\cite{DBLP:journals/siamcomp/KhotMS18}. This stronger structural input, together with the choice of degree scales below, removes a logarithmic loss from the copy bound. We adapt the notation from~\cite{pallavoor2022distance}.

We use $E$ for graph edges and $S\subseteq[n]$ for coordinate sets;
these play different roles in the path analysis.
Given a function $f$, let $E_f^-$ denote the set of violated edges:
\begin{equation}
\label{eq:set_of_violated_edges}
    E_f^-
    := \bigl\{
        \{x,y\} :
        x\preceq y,\ |y|-|x|=1,\ f(x)=1,\ f(y)=0
    \bigr\}.
\end{equation}
For a coloring $\mathrm{col}:E_f^-\to\{\mathrm{red},\mathrm{blue}\}$ that assigns each violated edge to either be red or blue, let $I^-_{f,\mathrm{col}}(x)$ count the number of red edges incident on $x$ if $f(x) = 1$ and the number of blue edges incident on $x$ if $f(x) = 0$. Each violated edge is then counted at exactly one endpoint. The following theorem from ~\cite{pallavoor2022distance} lower bounds the average of $\sqrt{I^-_{f,\mathrm{col}}(x)}$ in terms of the distance of $f$ from monotonicity over all colorings.

\begin{theorem}[{\cite[Theorem~2.9]{pallavoor2022distance}}]
\label{thm:robust-directed-isoperimetry}
There is an absolute constant $\kappa>0$ such that,
for every Boolean function $f:\{0,1\}^n\to\{0,1\}$
that is $\varepsilon$-far from monotone and every
coloring $\mathrm{col}$ of $E_f^-$ into red and blue,
\begin{equation}
\label{eq:robust-directed-isoperimetry}
    \mathbb E_{x\sim\mathrm{Unif}(\{0,1\}^n)}
    \left[\sqrt{I^-_{f,\mathrm{col}}(x)}\right]
    \geq \kappa\varepsilon.
\end{equation}
\end{theorem} 
Next, we use \Cref{thm:robust-directed-isoperimetry} to adapt the degree-removal argument of~\cite[Lemma~6.5]{DBLP:journals/siamcomp/KhotMS18}.

\begin{lemma}[Adapted from
{\cite[Lemma~6.5]{DBLP:journals/siamcomp/KhotMS18}}]
\label{lem:degree-selection}
There is an absolute constant $c>0$ such that the
following holds. Let $n\geq 1$, and suppose that
$f:\{0,1\}^n\to\{0,1\}$ is $\varepsilon$-far from
monotone, where $\varepsilon>0$.
Then there exist a power of two $1\leq d\leq n$,
a value $b\in\{0,1\}$, and a bipartite graph
$G=(A,B,E)$ with
\[
    A\subseteq f^{-1}(b),
    \qquad
    B\subseteq f^{-1}(1-b),
    \qquad
    E\subseteq E_f^-,
\]
such that
\[
    \deg_G(y)=d
    \quad\text{for every }y\in B,
    \qquad
    \deg_G(x)\leq 2d
    \quad\text{for every }x\in A,
\]

and, writing $\sigma=|B|/2^n$ for the uniform mass of $B$,
\begin{equation}
\label{eq:degree-selection}
    \sigma\geq c\varepsilon d^{-5/8}.
\end{equation}
\end{lemma}

\begin{proof}
Let $E_f^-$ be the edges of a bipartite graph with vertex classes $f^{-1}(1)$ and $f^{-1}(0)$. In particular, every edge joins an endpoint labeled $1$ to an endpoint labeled $0$. Let 
\begin{equation}
\label{eq:def_of_D_n}
\mathcal D_n :=\{1,2,4,\ldots,2^{\lfloor\log_2 n\rfloor}\}.
\end{equation}

We will process the values $d \in \mathcal D_n$ in decreasing order. For any $d$, let $U_d$ and $V_d$ be all sets of vertices of current degree at least $d$ with labels $1$ and $0$, respectively. Remove these vertices and all incident edges. 

Note that at the beginning of the $d$-th round, the maximum degree of every vertex remaining is less than $2d$. Furthermore, after the $d$-th round, all degrees are less than $d$ (and the property is always satisfied). After the last round with $d = 1$, no edges can remain. 

Next, for each edge in the graph, assign it to the vertex that was removed first, breaking ties arbitrarily. Color the edge red if it has been assigned to the lower (in coordinatewise partial order) vertex and blue otherwise. By definition, $I^-_{f,\mathrm{col}}(x)$ is exactly the number of edges assigned to $x$ under this coloring. Note that any vertex removed at scale $d$ has to be assigned fewer than $2d$ edges. We can write, using \Cref{eq:robust-directed-isoperimetry}, 
\[
    \kappa\varepsilon 2^{n}
    \leq
    \sum_{x\in\{0,1\}^n}
        \sqrt{I^-_{f,\mathrm{col}}(x)}
    \leq
    \sum_{d\in\mathcal D_n}
        (|U_d|+|V_d|)\sqrt {2d}.
\]
Here exactly $|U_d|+|V_d|$ vertices are removed at scale $d$, and each
such vertex is assigned fewer than $2d$ edges. The removal groups are disjoint;
vertices that are never removed are assigned no edges and contribute zero.

This implies that 
\begin{equation}
\label{eq:degree-selection-sum}
    \sum_{d\in\mathcal D_n}
        \frac{|U_d|+|V_d|}{2^n}\sqrt d
    \geq \frac{\kappa\varepsilon}{\sqrt{2}}.
\end{equation}

If $\max\{|U_d|,|V_d|\}<c\varepsilon 2^n d^{-5/8}$
for every $d\in\mathcal D_n$, then
\[
    \sum_{d\in\mathcal D_n}
        \frac{|U_d|+|V_d|}{2^n}\sqrt d
    <
    2c\varepsilon\sum_{j=0}^{\infty}2^{-j/8}
    =
    \frac{\kappa\varepsilon}{2\sqrt{2}},
\]
where $c=\kappa(1-2^{-1/8})/(4\sqrt{2})$.
This contradicts~\cref{eq:degree-selection-sum}.
Hence, for some $d\in\mathcal D_n$,
\[
    \max\{|U_d|,|V_d|\}
    \geq c\varepsilon 2^n d^{-5/8}.
\]

Now, let $d$ be the scale found above, and choose
$B\in\{U_d,V_d\}$ such that
\[
    |B|\geq c\varepsilon 2^n d^{-5/8}.
\]
Let $H$ be the current graph immediately before the
$d$-round, and let $A$ be the set of all neighbours
of $B$ in $H$. Thus $A$ lies on the opposite side
from $B$ but need not equal the other selected set
$U_d$ or $V_d$.

Every vertex in $B$ has degree at least $d$ in $H$.
For each $y\in B$, retain exactly $d$ incident edges,
and let $E$ be the set of all retained edges.
Since each edge has exactly one endpoint in $B$,
the resulting graph $G=(A,B,E)$ satisfies
\[
    \deg_G(y)=d
    \quad\text{for every }y\in B.
\]
Moreover, deleting edges cannot increase degrees, so
\[
    \deg_G(x)\leq\deg_H(x)<2d
    \quad\text{for every }x\in A.
\]
The set $B$ is unchanged, and hence
$\sigma=|B|/2^n\geq c\varepsilon d^{-5/8}$.
\end{proof}

Next, we establish the classical soundness bound needed for the quantum estimator. Recall that the total influence of a function $f$ is defined as
\[
    \mathbf I[f]
    :=\sum_{i=1}^n
    \Pr_{x\sim\mathrm{Unif}(\{0,1\}^n)}
        [f(x)\ne f(x^{\oplus i})],
\]
where $x^{\oplus i}$ denotes $x$ with its $i$-th coordinate flipped. As was done in \Cref{sec:fourier-upper}, we will use a coarse estimate of the total influence to handle the high-influence case separately. The classical lemma below therefore considers functions satisfying $\mathbf I[f]\le2\sqrt n$. We adapt the persistence and path-counting argument of ~\cite{DBLP:journals/siamcomp/KhotMS18}.

Following~\cite{DBLP:journals/siamcomp/KhotMS18}, we focus our analysis on violations of monotonicity in the middle layers of the Boolean hypercube. However, unlike ~\cite{DBLP:journals/siamcomp/KhotMS18}, we will use a band whose width depends on $\varepsilon$ in order to sharpen the complexity bound for our quantum tester. Let 
\begin{equation}
\label{eq:defn_of_w_and_mathcal_B}
    w:=\left\lceil
        \sqrt{\frac n2\log\frac{100}{\varepsilon}}
    \right\rceil,
    \qquad
    \mathcal B:=
    \left\{x\in\{0,1\}^n:
        \left||x|-\frac n2\right|\le w\right\}.
\end{equation}
We now describe a procedure for finding a violation in $\mathcal B$ whose success probability the quantum tester will estimate. 

Given an integer $t \geq 1$, sample $x\in\{0,1\}^n$ uniformly at random. Independently, choose upward or downward with equal probability. If the direction is upward, choose a uniformly random subset
\[
S\subseteq \{i\in[n]:x_i=0\}
\]
of size $t$, and define $y\in\{0,1\}^n$ by
\[
y_i=
\begin{cases}
1, & i\in S,\\
x_i, & i\notin S.
\end{cases}
\]
If the direction is downward, choose a uniformly random subset
\[
S\subseteq \{i\in[n]:x_i=1\}
\]
of size $t$, and define $y\in\{0,1\}^n$ by
\[
y_i=
\begin{cases}
0, & i\in S,\\
x_i, & i\notin S.
\end{cases}
\]
If fewer than $t$ coordinates are available in the chosen direction or if either $x$ or $y$ lies
outside $\mathcal B$, the procedure fails. Otherwise,
it succeeds if
\[
    (f(x),f(y))=
    \begin{cases}
        (1,0), & \text{upward},\\
        (0,1), & \text{downward}.
    \end{cases}
\]
We let $p_t(f)$ denote its success probability. The following lemma gives a lower bound on this probability for a suitable choice of $t$ when $f$ is $\varepsilon$-far from monotone and $\mathbf I[f]\le2\sqrt n$.

\begin{lemma}[Path soundness]
\label{lem:path-soundness}
There are absolute constants $c_0,c_1,n_0>0$ such that the
following holds. Let $n\ge n_0$ and

$0<\varepsilon\le1/2$. For each
\[
    d\in\mathcal D_n
    :=\{1,2,4,\ldots,2^{\lfloor\log_2 n\rfloor}\},
\]
define
\[
    t_d:=
    \max\left\{
        1,\left\lfloor
            c_0\varepsilon\sqrt n\,d^{-5/8}
        \right\rfloor
    \right\}.
\]
If $f:\{0,1\}^n\to\{0,1\}$ is $\varepsilon$-far from monotone
and $\mathbf I[f]\le2\sqrt n$, then for some
$d\in\mathcal D_n$,
\[
    p_{t_d}(f)\ge
    \frac{c_1\varepsilon t_d d^{3/8}}{n}.
\]
\end{lemma}

The proof is given in Appendix~\ref{app:path-soundness}.

\subsubsection{Quantum path estimator}
\label{sec:ordinary-path-observable}

We now construct the two measurements used by the path tester.
We keep the lengths $t_d$ from \Cref{lem:path-soundness} and assume
$n\geq n_0$ and $n^{-1/2}\leq\varepsilon\leq1/2$ throughout this subsection.
As in \Cref{sec:fourier-upper}, we estimate two terms on independent sets of
copies: the weighted disagreement between the endpoints, and a directional
correction. Their linear combination is an unbiased estimator for $p_{t_d}(f)$.

Recall from \eqref{eq:def_of_D_n} that
\[
    \mathcal D_n
    =\{1,2,4,\ldots,2^{\lfloor\log_2 n\rfloor}\}.
\]
Fix $d\in\mathcal D_n$, and let
\[
    \mathcal P_d:=
    \bigl\{(x,y)\in\mathcal B\times\mathcal B:
        x\preceq y,\ |y|-|x|=t_d\bigr\},
\]
where $\mathcal B$ is the Hamming band defined
in \eqref{eq:defn_of_w_and_mathcal_B}. Next, define
\[
    \kappa_d(x,y):=
    \begin{cases}
        \displaystyle\frac12\left(
            \binom{n-|x|}{t_d}^{-1}
            +\binom{|y|}{t_d}^{-1}
        \right),
        & (x,y)\in\mathcal P_d,\\[6pt]
        0, & \text{otherwise}.
    \end{cases}
\]
The two terms correspond to starting at $x$ and moving upward
or starting at $y$ and moving downward. Since each starting
vertex is chosen with probability $2^{-n}$, the probability
of selecting a pair $\{x,y\}$ is $2^{-n}\kappa_d(x,y)$. We may write
\begin{equation}
\label{eq:ordinary-path-probability-sum}
    p_{t_d}(f)
    =\frac1{2^n}\sum_{x,y}
        \kappa_d(x,y)f(x)(1-f(y)).
\end{equation}
Using the identity
\begin{equation}
\label{eq:ordinary-violation-decomposition}
    2f(x)(1-f(y))
    =\mathbf 1[f(x)\ne f(y)]+f(x)-f(y),
\end{equation}
we have that estimating $p_{t_d}(f)$ reduces to estimating the weighted disagreement between $f(x)$ and $f(y)$, together with the correction $f(x)-f(y)$.

Next, define 
\[
    q_d(f):=\frac1{2^n}\sum_{x,y}
        \kappa_d(x,y)\mathbf 1[f(x)\ne f(y)].
\]
For each $(x,y)\in\mathcal P_d$, let
\[
    \ket{\xi_{xy}}
    :=\frac{\ket{x}-\ket{y}}{\sqrt2}\otimes\ket{-},
\]
and define the observable
\begin{equation}
\label{eq:disagreement-observable}
    Q_d:=\sum_{(x,y)\in\mathcal P_d}
        \kappa_d(x,y)\ket{\xi_{xy}}\bra{\xi_{xy}}.
\end{equation}
For each $(x,y)\in\mathcal P_d$,
\[
    \bigl|\langle\xi_{xy}\mid\psi_f\rangle\bigr|^2
    =\frac{\bigl((-1)^{f(x)}-(-1)^{f(y)}\bigr)^2}{4\cdot2^n}
    =\frac{\mathbf 1[f(x)\ne f(y)]}{2^n}.
\]
Therefore,
\begin{equation}
\label{eq:disagreement-estimator-mean}
    \langle\psi_f|Q_d|\psi_f\rangle
    =\sum_{(x,y)\in\mathcal P_d}
        \kappa_d(x,y)
        \bigl|\langle\xi_{xy}\mid\psi_f\rangle\bigr|^2
    =q_d(f).
\end{equation}

For the second term, define
\[
    h_d(x):=\sum_y\kappa_d(x,y)-\sum_y\kappa_d(y,x),
    \qquad
    H_d:=\sum_x h_d(x)\ket{x}\bra{x},
\]
and let $Z:=\ket0\bra0-\ket1\bra1$ act on the qubit corresponding to the function value. Since each pair contributes equally to the two sums defining $h_d$, we have $\sum_x h_d(x)=0$. Therefore,
\[
\begin{aligned}
    \langle\psi_f|H_d\otimes Z|\psi_f\rangle
    &=\frac1{2^n}\sum_x h_d(x)(1-2f(x))\\
    &=-\frac2{2^n}\sum_x h_d(x)f(x)\\
    &=-\frac2{2^n}\sum_{x,y}
        \kappa_d(x,y)\bigl(f(x)-f(y)\bigr).
\end{aligned}
\]

Next, let $U_d$ be the outcome of measuring $Q_d$ on a copy of $\ket{\psi_f}$, and correspondingly, let $V_d$ be the outcome of measuring $H_d\otimes Z$ on an independent copy.
For an integer $m\geq1$, take $m$ independent samples of each outcome and define
\begin{equation}
\label{eq:ordinary-path-estimator}
    \widehat p_d
    :=\frac12\overline U_d-\frac14\overline V_d,
    \qquad
    \overline U_d:=\frac1m\sum_{j=1}^m U_d^{(j)},
    \qquad
    \overline V_d:=\frac1m\sum_{j=1}^m V_d^{(j)}.
\end{equation}
We have shown that $\mathbb E[\widehat p_d]=p_{t_d}(f)$. Next, we bound the variance. 

\begin{samepage}

\begin{lemma}[Variance of the path estimator]
\label{lem:path-estimator-variance}
Let $n\geq n_0$, $n^{-1/2}\leq\varepsilon\leq1/2$, and
$d\in\mathcal D_n$, with $t_d$ as in \Cref{lem:path-soundness}. Let the estimator in \Cref{eq:ordinary-path-estimator} use $m\geq1$ independent outcomes of each observable. There is a constant $C_{\mathrm{est}} > 0$ such that, whenever $\mathbf I[f]\leq2\sqrt n$, we have
\begin{equation}
\label{eq:ordinary-path-estimator-variance}
    \operatorname{Var}(\widehat p_d)
    \leq C_{\mathrm{est}} \frac{t_d}{m\sqrt n}.
\end{equation}
\end{lemma}

\end{samepage}

The proof is given in Appendix~\ref{app:path-variance}.

\subsubsection{Sample complexity}\label{sec:path-copy-complexity}

Finally, we are ready to bound the sample complexity of our procedure. Let's first construct a tester with error probability at most $\frac{1}{3}$, assuming $n^{-1/2}\leq\varepsilon\leq\frac{1}{2}$ and $n \geq n_0$.  As in \Cref{sec:fourier-upper}, we begin with a coarse influence test using $O(\sqrt{n})$ copies. In particular, we accept functions with $\mathbf I[f]\leq\sqrt n$ and reject functions with $\mathbf I[f]>2\sqrt n$ with error probability at most $\frac{1}{6}$. If we accept, we proceed as follows. 

For each $d\in\mathcal D_n$, where $\mathcal D_n$ is defined
in \Cref{eq:def_of_D_n}, choose
\[
    m_d:=
    \left\lceil
        \frac{64C_{\mathrm{est}}}{c_1^2}
        \frac{n^{3/2}}{\varepsilon^2t_d d^{3/4}}
    \right\rceil.
\]
We will use $2m_d$ copies of $\ket{\psi_f}$ in order to construct $\widehat p_d$, an estimate of $p_{t_d}(f)$. In particular, by Chebyshev's inequality and \Cref{eq:ordinary-path-estimator-variance}, we have 
\[
\begin{aligned}
    \Pr\!\left[
        |\widehat p_d-p_{t_d}(f)|
        \geq\frac{c_1\varepsilon t_d d^{3/8}}{4n}
    \right]
    &\leq
    \frac{16C_{\mathrm{est}}n^{3/2}}
         {c_1^2m_d\varepsilon^2t_d d^{3/4}}\\
    &\leq\frac{1}{4}.
\end{aligned}
\]
Next, let $r_d$ be the smallest odd integer that is at least $8 \log(12d)$. Using $m_d$ independent outcomes of each observable for each estimate, construct $r_d$ independent estimates of $\widehat p_d$, and let $\widetilde p_d$ be their median. By Hoeffding's Inequality, we have 
\[
    \Pr\!\left[
        |\widetilde p_d-p_{t_d}(f)|
        \geq\frac{c_1\varepsilon t_d d^{3/8}}{4n}
    \right]
    \leq e^{-r_d/8}
    \leq\frac{1}{12d}.
\]
The tester rejects if, for any $d\in\mathcal D_n$, we have
\[
    \widetilde p_d
    \geq\frac{c_1\varepsilon t_d d^{3/8}}{2n}.
\] A union bound tells us that the probability that at least one estimate of $\widetilde p_d$ differs from $p_{t_d}(f)$ by at least
$\frac{c_1\varepsilon t_d d^{3/8}}{4n}$ is at most 
\[
    \sum_{d\in\mathcal D_n}\frac{1}{12d}
    \leq\frac{1}{12}\sum_{j=0}^{\infty}2^{-j}
    =\frac{1}{6}.
\]

We may now check completeness and soundness. If $f$ is a monotone function, the $\mathbf I[f]\leq\sqrt n$ and
$p_{t_d}(f)=0$ for every $d$. The probability of error is at most $\frac{1}{6} + \frac{1}{6} = \frac{1}{3}$. Correspondingly, if $f$ is $\varepsilon$-far from monotone and $\mathbf I[f]\leq\sqrt n$, \cref{lem:path-soundness} tells us there exists a degree-$d$ such that 
\[
    p_{t_d}(f)\geq
    \frac{c_1\varepsilon t_d d^{3/8}}n,
\]
and our tester fails with probability at most $\frac{1}{6}$.

It remains to count the copies. Since
$\max\{1,\lfloor u\rfloor\}\geq u/2$ for $u\geq0$, the
definition of $t_d$ gives us
\[
    t_d\geq
    \frac{c_0}{2}\varepsilon\sqrt n\,d^{-5/8}.
\]
The number of copies used at degree $d$ is
\[
\begin{aligned}
    2m_dr_d
    &=O\!\left(
        \left(
            1+\frac{n^{3/2}}
                     {\varepsilon^2t_d d^{3/4}}
        \right)(1+\log d)
    \right)\\
    &=O\!\left(
        \frac n{\varepsilon^3}d^{-1/8}(1+\log d)
    \right),
\end{aligned}
\]
The total number of copies (including the coarse-influence tester) 
\[
\begin{aligned}
    O(\sqrt n)+\sum_{d\in\mathcal D_n}2m_dr_d
    &=O(\sqrt n)
      +O\!\left(
          \frac n{\varepsilon^3}
          \sum_{j=0}^{\lfloor\log_2 n\rfloor}
              (1+j)2^{-j/8}
      \right)\\
    &=O(n/\varepsilon^3),
\end{aligned}
\]
by convergence of the final sum. Finally, repeating the tester and taking a majority vote reduce the probability of error to $\delta$ via Hoeffding's Inequality. Therefore, the total number of copies scales as 
\[
    O\!\left(\frac n{\varepsilon^3}\log\frac1\delta\right).
\]

\section{Lower bound on monotonicity testing}\label{sec:two-sided-lower-bounds}

Next, we prove a lower bound on the number of quantum function states required to test monotonicity (with bounded two-sided error). We construct a pair of distributions over Boolean functions: a $\mathcal{D}_{\textit{yes}}$ distribution supported entirely on monotone functions, and a $\mathcal{D}_{\textit{no}}$ distribution supported on functions that, with high probability, are $\varepsilon$-far from monotone. We will show that any measurement that distinguishes the corresponding averaged $t$-copy ensembles with constant success probability requires at least
\[
t = \Omega\!\left(
\min\left\{
\frac{n}{\varepsilon^3},
\frac{n^{3/2}}{\varepsilon^2}
\right\}
\right)
\]
copies of the quantum function state.

\subsection{Block construction}
\label{subsec:block-construction}

Our construction combines many independently chosen local functions into a single global function. A similar idea has been used in classical query lower bounds for monotonicity ~\cite{belovs2016polynomial,chen2017talagrand} We place local functions in pairwise incomparable blocks of the Boolean cube in a way that monotonicity imposes no constraints between different local functions. We first construct the blocks and then we will specify the local functions used for the $\mathcal{D}_{\textit{yes}}$ and $\mathcal{D}_{\textit{no}}$ distributions. 

We separate the coordinates into two groups. If we write $n = d+m$, we may decompose each input as
\[
(y,x)\in\{0,1\}^d\times\{0,1\}^m,
\]
where $y$ selects the block and $x$ is the input to the function. In particular, if we let $U_1, \ldots U_R \subseteq \{0,1\}^d$ be pairwise disjoint blocks, then $y$ will determine which local function is applied, while $x$ will be an input to the function active on that block. Specifically, if $y \in U_j$, then 
\[
f(y, x) = f_j(x).
\]
Furthermore, we also require the blocks to be incomparable with each other -- if $j \neq k$, for all $y \in U_j$ and $y' \in U_k$, neither $y \leq y'$ nor $y' \leq y$. Finally, for $y \notin \bigcup_j U_j$, we fix $f(y,x)$ to be either $0$ or $1$ such that this value depends only on $y$ and not on $x$. These values are chosen such that the resulting global function $f$ is monotone whenever the local functions $f_j$ are monotone. 
As a small example, consider two selector coordinates and three local
coordinates, with $U_1=\{01\}$ and $U_2=\{10\}$. Let $f_1(x)=x_3$ in the first block and $f_2(x)=x_1$ in the second, and set the background values at $00$ and $11$ to $0$ and $1$, respectively. The resulting function is monotone: the two blocks are incomparable, and the background values are
compatible with both local functions.

Now, for each block $U_j$, we will define its relative mass by
\begin{equation}
\label{eq:p_j_mass}
p_j := \frac{|U_j|}{2^d}.
\end{equation}
We will also use
\begin{equation}
\label{eq:block-mass-parameters}
\beta := \sum_{j=1}^R p_j,
\qquad
p_{\max} := \max_{1\leq j\leq R} p_j.
\end{equation}
Therefore, $\beta$ represents the total mass of the blocks, while $p_{\max}$ will be the mass of the largest individual block. The following lemma allows independent local choices to contribute a constant total distance while each block remains difficult to observe.
\begin{samepage}
\begin{lemma}[Selector blocks]
\label{lem:selector-blocks}
For sufficiently large $d$, there exist pairwise disjoint and
incomparable selector blocks satisfying
\begin{equation}
\label{eq:bound_on_beta_and_pmax}
\beta \geq \frac{1}{64},
\qquad
p_{\max}\leq 2^{-r},
\qquad
r:=\left\lfloor\frac{\sqrt d}{10}\right\rfloor.
\end{equation}
Here $\beta$ and $p_{\max}$ are defined in \eqref{eq:block-mass-parameters}.
\end{lemma}
\end{samepage}

The proof is given in Appendix~\ref{app:selector-mass}.

\subsection{The local functions}\label{subsec:local-functions}

We now define the allowed local functions that can be assigned to the blocks $U_j$. Each local function will be defined on all inputs $x \in \{0,1\}^m$. Furthermore, within this local input space, we will choose a central band of Hamming-weight layers. The local functions will differ only within this band, taking the value $0$ below it and $1$ above it. The width of the band will allow us to control the distance from monotonicity and the distinguishability of the resulting quantum ensembles. 

Choose integers $0\leq L<U\leq m$ with $L+U=m$, and define
\[
B:=\{x\in\{0,1\}^m:L\leq |x|\leq U\}.
\]
Let $w:=U-L$ be the width of the band. For each $i\in[m]$, define $f_i^+$ and $f_i^-$ by setting
\[
f_i^+(x)=x_i,
\qquad
f_i^-(x)=1-x_i
\qquad\text{for }x\in B.
\] 
Furthermore, define two functions that are constant inside the band:
\[
h_0(x)=0,
\qquad
h_1(x)=1
\qquad\text{for }x\in B.
\]

For all of these functions, set the value to $0$ when $|x|<L$ and to $1$ when $|x|>U$. Note that $h_0$, $h_1$, and $f_i^+$ are monotone. The distance of $f_i^-$ from the set of monotone functions is exactly
\begin{equation}
\label{eq:distance_from_monotone_functions}
\delta_w:=
\frac12\Pr_{x_{-i}}
\left[L\leq |x_{-i}|\leq U-1\right],
\end{equation}
where $x_{-i}$ denotes the input with coordinate $i$ removed and is uniform on $\{0,1\}^{m-1}$. Each $x_{-i}$ within this range gives a violated edge where both endpoints are within the band. Since these edges are disjoint, making $f_i^-$ monotone requires
changing at least one endpoint of each edge, and hence at least a $\delta_w$ fraction of all inputs. Conversely, changing the value from $0$ to $1$ at the endpoint with $x_i=1$ on each such edge produces the monotone function
\[
\mathbf{1}\{|x_{-i}|\geq L\}.
\]
This changes exactly $\delta_w$ fraction of the inputs, proving the bound. 

\subsection{Global distributions}\label{subsec:global-distributions}
We now fix one choice of the selector terms and blocks
from Appendix~\ref{app:selector-mass} satisfying
\Cref{eq:bound_on_beta_and_pmax}. The same blocks and background values will be used by both distributions. Only their local functions will be randomized. We define $\mathcal{D}_{\textit{yes}}$ and
$\mathcal{D}_{\textit{no}}$ by choosing a local function
independently for each block $U_j$. Let $0<\rho\leq\alpha\leq1$ be parameters that will be chosen later (depending on the $\varepsilon$-regime). Let the probabilities of selecting
each local function be
\begin{table}[htbp]
\[
\begin{array}{c|cc}
\text{Local function}
& \mathcal{D}_{\textit{yes}}
& \mathcal{D}_{\textit{no}} \\ \hline
f_i^+ \quad (i\in[m]) & \alpha/m & (\alpha-\rho)/m \\
f_i^- \quad (i\in[m]) & 0 & \rho/m \\
h_0 & (1-\alpha)/2 & (1-\alpha)/2 \\
h_1 & (1-\alpha)/2 & (1-\alpha)/2
\end{array}
\]
\caption{Distributions over local functions.}
\label{tab:local-distributions}
\end{table}
Since $\mathcal{D}_{\textit{yes}}$ uses
only monotone local functions, every function in this distribution is monotone. For $\mathcal{D}_{\textit{no}}$, we will show that, with high probability, every function is at least $\varepsilon$-far for monotone. 

Let $f\sim\mathcal{D}_{\textit{no}}$, and let $Z_j$ be the variable that indicates that the block $U_j$ was assigned a reversed dictator in the construction of $f$. Then, $Z_j$ are independent Bernoulli variables with parameter $\rho$. Since each reversed dictator function has distance $\delta_w$
from a monotone function (see \Cref{eq:distance_from_monotone_functions})
and $p_j$ represents the weight of block $U_j$
(see \Cref{eq:p_j_mass}), we have
\begin{equation}
\label{eq:global-distance-from-monotonicity}
\operatorname{dist}(f,\rev{\mathrm{MON}}_{d+m})
=
\delta_w\sum_{j=1}^R p_jZ_j.
\end{equation}
This bound can be attained by repairing each block separately. 

Next, let $Z:=\sum_j p_jZ_j$. We have 
\begin{equation}
\label{eq:expected-negative-mass}
\mathbb{E}[Z]
= \mathbb{E}\left[\sum_j p_j Z_j\right]
= \sum_j p_j \mathbb{E}[Z_j]
= \rho \cdot \beta,
\end{equation}
and, by independence,
\[
\operatorname{Var}(Z)
= \rho(1-\rho)\sum_j p_j^2
\leq \rho\sum_j p_j^2
\leq \rho \cdot p_{\max}\sum_j p_j
= \rho \cdot p_{\max} \cdot \beta.
\]
Assume we choose the parameters such that $\delta_w\rho\beta\geq 2\varepsilon$. Since $\operatorname{dist}(f,\rev{\mathrm{MON}}_{d+m})=\delta_w Z$,
\[ 
\Pr[\operatorname{dist}(f,\rev{\mathrm{MON}}_{d+m})<\varepsilon] = \Pr\left[Z<\frac{\varepsilon}{\delta_w}\right] \leq\Pr\left[Z<\frac{\rho\beta}{2}\right]. 
\]
Applying Chebyshev's inequality (using $\mathbb E[Z]=\rho\beta$), we get 
\begin{equation}
\label{eq:no-farness-chebyshev}
\begin{aligned} 
\Pr\left[Z<\frac{\rho\beta}{2}\right] 
&\leq \Pr\left[|Z-\mathbb E[Z]|\geq\frac{\rho\beta}{2}\right]
&\leq \frac{4p_{\max}}{\rho\beta}. 
\end{aligned} 
\end{equation}
We will later choose parameters such that this probability tends to 0. Lastly, note that if $\alpha = \rho = 1$, every function in $\mathcal{D}_{\textit{no}}$ is at least $\beta\delta_w\geq 2\varepsilon$-far from monotone. 

\subsection{Poissonization}\label{subsec:coherent-poissonization}

To prove the lower bound, we will start by considering a coherent superposition over different numbers of copies. This will allow us to represent the states using a tensor product across the blocks $U_1, \ldots U_R$, making it easier to compare the states arising from $\mathcal{D}_{\textit{yes}}$ and $\mathcal{D}_{\textit{no}}$.

For $\lambda>0$, define
\begin{equation}
\label{eq:poissonization_construction}
\ket{\Phi_\lambda(f)}
:=
e^{-\lambda/2}
\bigoplus_{k=0}^{\infty}
\frac{\lambda^{k/2}}{\sqrt{k!}}\,
\ket{\psi_f}^{\otimes k},
\qquad
\ket{\psi_f}^{\otimes 0}:=\ket{\emptyset}.
\end{equation}
where $\ket{\emptyset}$ is simply a fixed unit vector independent of $\ket{\psi_f}$, representing $k=0$ copies. For each $k \geq 0$, let $\Pi_k$ represent the projector:
\begin{equation}
\label{eq:k-copy-measurement}
\Pi_k
=
0\oplus\cdots\oplus
I_{\mathcal H^{\otimes k}}
\oplus\cdots,
\end{equation}
where $\mathcal H$ is the Hilbert space for a single $\rev{\ket{\psi_f}}$. The projective measurement $\{\Pi_k\}_{k \geq 0}$ identifies the number of copies. Applied to $\ket{\Phi_\lambda(f)}$, the measurement gives an outcome $k$ with probability
\[
\bigl\|\Pi_k\ket{\Phi_\lambda(f)}\bigr\|^2
=
e^{-\lambda}\frac{\lambda^k}{k!}.
\]
Furthermore, the normalized state will be $\ket{\psi_f}^{\otimes k}$. We can think of the measurement as producing an outcome $K\sim\operatorname{Poisson}(\lambda)$ where, conditioned on $K = k$, the state is $\ket{\psi_f}^{\otimes k}$. We will eventually show how a bound for the coherent superposition states will imply a lower bound for testers that are given a fixed number of copies.

The selector blocks remain fixed as in \Cref{subsec:global-distributions};
in particular, $f$ and $g$ below share both these blocks and their background.
Next, we will show why we work with the coherent superposition. Consider two global functions $f$ and $g$ obtained from our block construction, with local functions $f_j$ and $g_j$ respectively, assigned to each block $U_j$. The functions agree outside of the blocks. The distance between the functions may be written as 
\begin{equation}
\operatorname{dist}(f,g)
=
\sum_{j=1}^R
p_j\operatorname{dist}(f_j,g_j).
\label{eq:block-distance-decomposition}
\end{equation}
Furthermore, recall that the overlap between their quantum function state representation may be written as  
\[
\langle\psi_f\mid\psi_g\rangle
=
2^{-n}
\sum_{z\in\{0,1\}^n}\mathbf{1}\{f(z)=g(z)\}
=
1-\operatorname{dist}(f,g).
\]
For $t$ copies of the quantum function states, the overlap may be written as
\begin{align*}
\left\langle
\psi_f^{\otimes t}
\middle|
\psi_g^{\otimes t}
\right\rangle
&=
\prod_{\ell=1}^{t}
\langle\psi_f\mid\psi_g\rangle\\
&=
\left(
1-\sum_{j=1}^R
p_j\operatorname{dist}(f_j,g_j)
\right)^t.
\end{align*}
This product in general does not factor over the blocks. 

As we will see, the use of the coherent superposition states will turn the overlap into an exponential which does factor over blocks. Specifically, we have 
\begin{equation}
\label{eq:coherent-overlap}
\begin{aligned}
\langle\Phi_\lambda(f)\mid\Phi_\lambda(g)\rangle
&=
e^{-\lambda}
\sum_{k=0}^{\infty}
\frac{\lambda^k}{k!}
\left\langle
\psi_f^{\otimes k}
\middle|
\psi_g^{\otimes k}
\right\rangle
\\
&=
e^{-\lambda}
\sum_{k=0}^{\infty}
\frac{\lambda^k}{k!}
\bigl(\langle\psi_f\mid\psi_g\rangle\bigr)^k
\\
&=
\exp\!\left(
-\lambda+\lambda\langle\psi_f\mid\psi_g\rangle
\right)
\\
&=
e^{-\lambda\operatorname{dist}(f,g)}.
\end{aligned}
\end{equation}
where we used that the overlap of the $k=0$ terms is $1$. Substituting \Cref{eq:block-distance-decomposition} gives
\begin{align*}
\langle\Phi_\lambda(f)\mid\Phi_\lambda(g)\rangle
&=
\exp\!\left(
-\lambda\sum_{j=1}^R
p_j\operatorname{dist}(f_j,g_j)
\right)\\
&=
\prod_{j=1}^R
e^{-\lambda p_j\operatorname{dist}(f_j,g_j)}.
\end{align*}
Therefore, the overlap is written as product over the individual blocks depending only on the distance between the local functions $f_j$ and $g_j$. In the next subsection, we will see how to use this factorization to represent the global coherent superposition state as a product of local coherent states, one for each block.

\subsection{Factorization across blocks}
\label{subsec:block-factorization}
Let
\[
\mathcal F
:=
\operatorname{supp}(\mathcal D_{\textit{yes}})
\cup
\operatorname{supp}(\mathcal D_{\textit{no}}).
\]
All functions in $\mathcal F$ come from the block
construction described earlier and share the same fixed blocks $U_1, \ldots U_R \subseteq \{0,1\}^d$. Each $f \in \mathcal F$ is specified by the local function $f_j:\{0,1\}^m \to \{0,1\}$ assigned to each of these blocks, where $d+m = n$. In particular, for each $j\in[R]$,
\[
f(y,x)=f_j(x)
\qquad
\text{for every }y\in U_j
\text{ and }x\in\{0,1\}^m.
\]
Additionally, we have $f(y,x) = 0$ if $y$ satisfies none of $T_j$, and $f(y,x) = 1$ when $y$ satisfies at least two of them. These values depend only on $y$ and are the same across all $f \in \mathcal{F}$. 

Next, for any function $h:\{0,1\}^m \to \{0,1\}$, we may write 
\[
\ket{\psi_h}
=
2^{-m/2}
\sum_{x\in\{0,1\}^m}\ket{x}\ket{h(x)},
\]
and for any $\mu > 0$, we set 
\[
\ket{\Phi_\mu(h)}
:=
e^{-\mu/2}
\bigoplus_{k=0}^{\infty}
\frac{\mu^{k/2}}{\sqrt{k!}}\,
\ket{\psi_h}^{\otimes k}
\qquad
\rev{\ket{\psi_h}^{\otimes 0}}:=\ket{\emptyset_{\mathrm{local}}}.
\]
Furthermore,  $\ket{\Phi_0(h)}$ is also taken to be the unit vector $\ket{\emptyset_{\mathrm{local}}}$. This is the same coherent superposition construction as \Cref{eq:poissonization_construction}. Furthermore, analogous to \Cref{eq:coherent-overlap}, we have 
\begin{equation}
\label{eq:overlap_between_phi}
\langle\Phi_\mu(h)\mid\Phi_\mu(h')\rangle
=
e^{-\mu\operatorname{dist}(h,h')}.
\end{equation}

We will now construct an isometry that maps each global coherent superposition state $\ket{\Phi_\lambda(f)}$ to a tensor product of coherent states (with one factor per block). We will show how the trace distance between the averaged global coherent states (over $\mathcal{D}_{\textit{yes}}$ and $\mathcal{D}_{\textit{no}}$) is equal to the trace distance between the corresponding tensor products of the averaged coherent states for individual blocks.

For each $f\in\mathcal F$, we can define the product state
\begin{equation}
\begin{aligned}
\ket{\chi_\lambda(f)}
&:=
\bigotimes_{j=1}^R
\ket{\Phi_{\lambda p_j}(f_j)}\\
&=
\ket{\Phi_{\lambda p_1}(f_1)}
\otimes\cdots\otimes
\ket{\Phi_{\lambda p_R}(f_R)}.
\end{aligned}
\label{eq:local-coherent-product}
\end{equation}
This tensor product has one factor for each block.
The $j$-th factor uses the local function $f_j$
and Poisson parameter $\lambda p_j$. If a block is
empty, then $p_j=0$ and its factor is the fixed
vacuum state($\ket{\emptyset_{\mathrm{local}}}$), independent of the local function.

For $b\in\{\textit{yes},\textit{no}\}$, let $\nu_b$
denote the distribution over local functions given
by the probabilities in the $\mathcal D_b$ column
of Table~\ref{tab:local-distributions}.  Sampling a function from $\mathcal D_b$ is equivalent to sampling $f_1,\ldots,f_R$ independently from $\nu_b$ and assembling them using the fixed
block construction.

Define the averaged global coherent state by
\[
\Omega_b(\lambda)
:=
\mathbb E_{f\sim\mathcal D_b}
\left[
\ket{\Phi_\lambda(f)}\bra{\Phi_\lambda(f)}
\right],
\]
and define the averaged local
coherent state by
\[
\sigma_b(\mu)
:=
\mathbb E_{h\sim\nu_b}
\left[
\ket{\Phi_\mu(h)}\bra{\Phi_\mu(h)}
\right].
\]

We use $\Omega_b(\lambda)$ for the averaged global coherent state,
$\sigma_b(\mu)$ for one averaged local coherent state, and
$\rho_b^{(t)}$ below for the averaged state of a fixed $t$ copies. The distributions $\nu_b$ select local functions, and $\mathcal D_b$ select global functions. The parameter in block $j$ is always $\mu=\lambda p_j$.

\begin{lemma}[Block factorization]
\label{lem:block-factorization}
There is an isometry from the span of the global coherent states to
the span of the corresponding product states satisfying
\begin{equation}
V_\lambda\ket{\Phi_\lambda(f)}
=
\ket{\chi_\lambda(f)}
=
\bigotimes_{j=1}^R
\ket{\Phi_{\lambda p_j}(f_j)}
\qquad\text{for every }f\in\mathcal F.
\label{eq:coherent-block-isometry}
\end{equation}
For either local distribution, the averaged state satisfies
\[
V_\lambda\Omega_b(\lambda)V_\lambda^\dagger
=\bigotimes_{j=1}^R\sigma_b(\lambda p_j),
\qquad b\in\{\textit{yes},\textit{no}\}.
\]
Consequently,
\begin{equation}
\left\|
\Omega_{\textit{yes}}(\lambda)
-
\Omega_{\textit{no}}(\lambda)
\right\|_1
=
\left\|
\bigotimes_{j=1}^R \sigma_{\textit{yes}}(\lambda p_j)
-
\bigotimes_{j=1}^R \sigma_{\textit{no}}(\lambda p_j)
\right\|_1.
\label{eq:coherent-product-distance}
\end{equation}
\end{lemma}

The global coherent states and the product states have identical
pairwise inner products, so the correspondence extends to an isometry.
Independence of the local choices then factors the averaged states.
The complete proof of \Cref{lem:block-factorization} is in
Appendix~\ref{app:block-isometry}.

\subsection{Reduction to a fixed number of copies}
\label{subsec:fixed-copy-reduction}

Next, we will show how to use the trace distance between the averaged coherent ensembles to bound the trace distance between averaged $t$-copy ensembles. For $b\in\{\textit{yes},\textit{no}\}$ and an integer
$t\geq1$, define
\[
\rho_b^{(t)}
:=
\mathbb E_{f\sim\mathcal D_b}
\left[
\bigl(\ket{\psi_f}\bra{\psi_f}\bigr)^{\otimes t}
\right].
\] 

Starting from the coherent superposition state
$\ket{\Phi_\lambda(f)}$, perform the copy-number
measurement $\{\Pi_k\}_{k\geq0}$ defined in
\Cref{eq:k-copy-measurement}. This gives an outcome
$K\sim\operatorname{Poisson}(\lambda)$ and leaves
$K$ copies of $\ket{\psi_f}$. If $K \geq t$, retain $t$ copies of the state and discard the \rev{rest}. If $K < t$, discard the state and output a failure flag state $\ket{\mathrm{fail}}$, orthogonal to the \rev{$t$-copy space}. This procedure is independent of our function $f$ and defines a quantum channel, which we will call $\mathcal E_t$. If we let 
\[
q_{\lambda,t}
:=
\Pr_{K\sim\operatorname{Poisson}(\lambda)}[K\geq t],
\]
we can write
\[
\mathcal E_t\bigl(\Omega_b(\lambda)\bigr)
=
q_{\lambda,t}\rho_b^{(t)}
\oplus
(1-q_{\lambda,t})
\ket{\mathrm{fail}}\bra{\mathrm{fail}}
\]
for $b\in\{\textit{yes},\textit{no}\}$. Since quantum channels cannot increase the trace distance, we may write 
\begin{equation}
\begin{aligned}
q_{\lambda,t}
\left\|
\rho_{\textit{yes}}^{(t)}
-
\rho_{\textit{no}}^{(t)}
\right\|_1
&=
\left\|
\mathcal E_t\bigl(\Omega_{\textit{yes}}(\lambda)\bigr)
-
\mathcal E_t\bigl(\Omega_{\textit{no}}(\lambda)\bigr)
\right\|_1\\
&\leq
\left\|
\Omega_{\textit{yes}}(\lambda)
-
\Omega_{\textit{no}}(\lambda)
\right\|_1.
\end{aligned}
\label{eq:poisson-fixed-copy-comparison}
\end{equation}

Next, we can choose $\lambda = 2t$ to ensure that the procedure succeeds with probability at least $1/2$. If $K\sim\operatorname{Poisson}(2t)$, we have
$\mathbb E[K]=\operatorname{Var}(K)=2t$. By Chebyshev's inequality, we get
\[
\Pr[K<t]
\leq
\Pr[|K-2t|\geq t+1]
\leq
\frac{2t}{(t+1)^2}
\leq
\frac12.
\]
Since $q_{2t,t}\geq1/2$, applying
\Cref{eq:poisson-fixed-copy-comparison} and
\Cref{eq:coherent-product-distance} yields
\begin{align}
\left\|
\rho_{\textit{yes}}^{(t)}
-
\rho_{\textit{no}}^{(t)}
\right\|_1
&\leq
2\left\|
\Omega_{\textit{yes}}(2t)
-
\Omega_{\textit{no}}(2t)
\right\|_1
\\
&=
2\left\|
\bigotimes_{j=1}^R
\sigma_{\textit{yes}}(2tp_j)
-
\bigotimes_{j=1}^R
\sigma_{\textit{no}}(2tp_j)
\right\|_1.
\end{align}

In the next subsection, we will bound the right-hand side
by comparing the local averaged coherent states.

\subsection{Local fidelities}
\label{subsec:local-coherent-comparison}

We will bound the trace distance between the averaged states
by lower bounding their fidelity. For density operators
$\omega$ and $\tau$, let
\[
F(\omega,\tau)
:=
\left\|\sqrt{\omega}\sqrt{\tau}\right\|_1.
\]
The Fuchs--van de Graaf inequality~\cite{fuchs1999cryptographic} says that
\[
\|\omega-\tau\|_1
\leq
2\sqrt{1-F(\omega,\tau)^2}.
\]
\rev{Define the root fidelity between the local averaged coherent states by}
\[
F_\mu
:=
F\bigl(
\sigma_{\textit{yes}}(\mu),
\sigma_{\textit{no}}(\mu)
\bigr).
\]
Since every block uses the same local ensembles, the function $F_\mu$ is independent of the block index $j$. The fidelity of the $j$-th block can be obtained by evaluating the function at $\mu = 2tp_j$.

\begin{samepage}
\begin{revision}
\begin{lemma}[Comparison through local fidelities]
\label{lem:local-to-global-fidelity}
For the fixed selector blocks and independent local choices above,
the averaged $t$-copy ensembles satisfy, for every integer $t\geq1$,
\[
    \bigl\|\rho_{\textit{yes}}^{(t)}
              -\rho_{\textit{no}}^{(t)}\bigr\|_1^2
    \leq32\sum_{j=1}^R(1-F_{2tp_j}).
\]
\end{lemma}
\end{revision}
\end{samepage}
\begin{proof}
We then have that
\begin{equation}
\label{eq:rho_yes_and_no_t_copies}
\begin{aligned}
\left\|
\rho_{\textit{yes}}^{(t)}
-
\rho_{\textit{no}}^{(t)}
\right\|_1
&\leq
2\left\|
\Omega_{\textit{yes}}(2t)
-
\Omega_{\textit{no}}(2t)
\right\|_1
\\
&\leq
4\sqrt{
1-
F\bigl(
\Omega_{\textit{yes}}(2t),
\Omega_{\textit{no}}(2t)
\bigr)^2
}
\\
&=
4\sqrt{
1-
\prod_{j=1}^{R}
F\bigl(
\sigma_{\textit{yes}}(2tp_j),
\sigma_{\textit{no}}(2tp_j)
\bigr)^2
}.
\end{aligned}
\end{equation}
Using \cref{eq:rho_yes_and_no_t_copies}, we may write 
\begin{align*}
    \bigl\|\rho_{\textit{yes}}^{(t)}
              -\rho_{\textit{no}}^{(t)}\bigr\|_1^2
    &\leq16\left(1-\prod_{j=1}^R F_{2tp_j}^2\right)\\
    &\leq32\sum_{j=1}^R(1-F_{2tp_j}).
\end{align*}
since $0 \leq F_{2tp_j} \leq 1$. 

\end{proof}

\subsection{Bounding the local states}
\label{subsec:local-fidelity-estimate}

Using the local functions $f_i^+$, $f_i^-$, $h_0$, and $h_1$ defined in \Cref{tab:local-distributions}, we can write the local averaged states as:
\begin{align*}
\sigma_{\textit{yes}}(\mu)
&=
\frac{\alpha}{m}
\sum_{i=1}^m
\ket{\Phi_\mu(f_i^+)}\!\bra{\Phi_\mu(f_i^+)}
+
\frac{1-\alpha}{2}
\sum_{s=0}^1
\ket{\Phi_\mu(h_s)}\!\bra{\Phi_\mu(h_s)},
\\ \\
\sigma_{\textit{no}}(\mu)
&=
\frac{\alpha-\rho}{m}
\sum_{i=1}^m
\ket{\Phi_\mu(f_i^+)}\!\bra{\Phi_\mu(f_i^+)}
+
\frac{\rho}{m}
\sum_{i=1}^m
\ket{\Phi_\mu(f_i^-)}\!\bra{\Phi_\mu(f_i^-)}
\\
&\qquad \qquad{}+
\frac{1-\alpha}{2}
\sum_{s=0}^1
\ket{\Phi_\mu(h_s)}\!\bra{\Phi_\mu(h_s)}.
\end{align*}
We also have that $m \geq 2$ and $0\leq\rho\leq\alpha\leq1$. Furthermore, the local functions agree outside the band
\[
B=\{x\in\{0,1\}^m:L\leq |x|\leq U\},
\]
where $L$ and $U$ are cutoffs satisfying $0\leq L\leq U\leq m$ and $L+U=m$. 

Let
\begin{equation}
\label{eq:band-mass}
p:=\frac{|B|}{2^m}.
\end{equation}
and define 
\[
    \ket{z_g}
    :=
    \frac{1}{\sqrt{|B|}}
    \sum_{x\in B}(-1)^{g(x)}\ket{x}
\]
for each local function $g$. Since any local functions $g, h$ agree outside the band $B$, we may write 
\begin{equation}
\label{eq:band-sign-overlap}
    \langle z_g\mid z_h\rangle
    =
    \frac{1}{|B|}
    \sum_{x\in B}(-1)^{g(x)+h(x)}
    =
    1-\frac{2\operatorname{dist}(g,h)}{p}.
\end{equation}
Furthermore, let 
\[
    \ket{\bar z}
    :=
    \frac{1}{m}\sum_{i=1}^m\ket{z_{f_i^+}},
    \qquad
    C:=
    \frac{\alpha}{m}\sum_{i=1}^m
    \ket{z_{f_i^+}}\bra{z_{f_i^+}}
    +(1-\alpha)\ket{z_{h_0}}\bra{z_{h_0}}.
\]
Here, $\ket{\bar z}$ is the average of the vectors
$\ket{z_{f_i^+}}$, and $C$ is the average of the projectors
$\ket{z_g}\bra{z_g}$ under either distribution
$\nu_{\textit{yes}}$ or $\nu_{\textit{no}}$. 
\begin{samepage}
\begin{lemma}[\rev{Local fidelity bound}]
\label{lem:local-fidelity-deficit}
\rev{For the local ensembles and band parameters defined above, if
$0 <\rho\leq\alpha\leq1$, then
\[
1-F_\mu
\leq
\frac{\rho^2}{1-\alpha+\rho}\,
\mu p\|\bar z\|^2
+\frac{\mu^2p^2}{4}.
\]
If $\rho = 0$, the local distributions coincide and $F_{\mu} = 1$.}
\end{lemma}
\end{samepage}

The proof is given in Appendix~\ref{app:local-fidelity-deficit}. \\

Applying \cref{eq:local-fidelity-deficit} with $\mu=2tp_j$
for each j gives
\begin{equation}
\label{eq:general-ensemble-distance}
\begin{aligned}
\bigl\|\rho_{\textit{yes}}^{(t)}
      -\rho_{\textit{no}}^{(t)}\bigr\|_1^2
&\leq32\sum_{j=1}^R(1-F_{2tp_j})\\
&\leq
\frac{64\rho^2}{1-\alpha+\rho}\,
tp\|\bar z\|^2\sum_{j=1}^R p_j
+32t^2p^2\sum_{j=1}^R p_j^2\\
&\leq
\frac{64\rho^2}{1-\alpha+\rho}\,
t\beta p\|\bar z\|^2
+32t^2p^2\beta p_{\max}.
\end{aligned}
\end{equation}
The last inequality uses the definitions in
\cref{eq:block-mass-parameters} and
$\sum_jp_j^2\leq p_{\max}\sum_jp_j=\beta p_{\max}$.

\subsection{Parameter choices and the copy lower bound}
\label{sec:parameters-copy-lower-bound}

We may now choose the parameters so that a function sampled from $\mathcal D_{\textit{no}}$ is $\varepsilon$-far from monotone with high probability, while the averaged $t$-copy states remain difficult to distinguish. Recall that $w=U-L$ is the width of the central band on the $m$ action coordinates. Since $L+U=m$, the band is
\[
B=
\left\{
x\in\{0,1\}^m:
\frac{m-w}{2}\leq |x|\leq\frac{m+w}{2}
\right\}.
\]
The band mass $p$, as defined in \Cref{eq:band-mass}, is the fraction of inputs lying in $B$. Furthermore, the quantity $\delta_w$, defined in \cref{eq:distance_from_monotone_functions}, is the distance from monotonicity for each $f_{i}^{-}$. We have
\[
p=\frac{|B|}{2^m}
  =\frac1{2^m}\sum_{k=L}^{U}\binom{m}{k},
\qquad
\delta_w=\frac1{2^m}\sum_{k=L}^{U-1}\binom{m-1}{k}.
\]
Only blocks assigned a function $f_i^-$ contribute to the
distance from monotonicity. Each block receives such a function with probability $\rho$, so \cref{eq:global-distance-from-monotonicity,eq:expected-negative-mass} give an expected distance of $\beta\rho\delta_w$. For fixed $m$, choosing $w$ determines $p$, $\delta_w$, and $\|\bar z\|^2$. We will choose $w$ and $\rho$ to ensure that the functions are far with high probability and then apply \cref{eq:general-ensemble-distance}
to bound the distinguishability of the ensembles.

Next, we take $m$ and $w$ odd so that the endpoints $L=(m-w)/2$ and $U=(m+w)/2$ are integers and so the width-one band used below is admissible. Using Pascal's identity, we have
\[
\begin{aligned}
p
&=\frac1{2^m}\sum_{k=L}^{U}
  \left[\binom{m-1}{k}+\binom{m-1}{k-1}\right]\\
&=2\delta_w+\frac{2}{2^m}\binom{m-1}{L-1},
\end{aligned}
\]
where we use the convention that $\binom{m-1}{-1}=\binom{m-1}{m}=0$. Since $L+U=m$ and $U-L\geq1$, we have $L\leq(m-1)/2$. Therefore,
\[
\binom{m-1}{L-1}
\leq \binom{m-1}{L}
\leq \sum_{k=L}^{U-1}\binom{m-1}{k}
=2^m\delta_w,
\]
and we get that
\begin{equation}
\label{eq:band-mass-distance-comparison}
2\delta_w\leq p\leq4\delta_w.
\end{equation}

Stirling's formula tells us that
\[
2^{-m}\binom{m}{k}=\Theta(m^{-1/2})
\qquad\text{for } |k-m/2|\leq\sqrt m.
\]
If $w\leq\sqrt m$, the band contains $w+1$ layers satisfying this constraint and we get $p=\Theta((w+1)/\sqrt m)$. If $w>\sqrt m$, the band contains at least $\sqrt m$ central layers, each of mass $\Theta(m^{-1/2})$. Therefore, $p=\Theta(1)$ since $p$ is at most $1$. Using \cref{eq:band-mass-distance-comparison},
we get
\begin{equation}
\label{eq:band-width-estimates}
\Theta(\delta_w) = p
=\Theta\!\left(\min\left\{1,\frac{w+1}{\sqrt m}\right\}\right).
\end{equation}
Furthermore,
\[
\delta_1
=\frac1{2^m}\binom{m-1}{(m-1)/2}
=\Theta(m^{-1/2}),
\qquad
\delta_m=\frac12.
\]

Next, we bound $\|\bar z\|^2$. Recall that
$|\bar z\rangle$ is the average of the vectors
corresponding to the functions $f_i^+$. Since
$f_i^+(x)=x_i$ on $B$,
\[
\begin{aligned}
|\bar z\rangle
&=\frac1m\sum_{i=1}^m |z_{f_i^+}\rangle\\
&=\frac1{\sqrt{|B|}}\sum_{x\in B}
  \left(\frac1m\sum_{i=1}^m(-1)^{x_i}\right)|x\rangle\\
&=\frac1{\sqrt{|B|}}\sum_{x\in B}
  \frac{m-2|x|}{m}|x\rangle.
\end{aligned}
\]
Using $p=|B|/2^m$, we get
\begin{equation}
\label{eq:band-moment-identity}
p\|\bar z\|^2
=\frac1{2^m}\sum_{x\in B}
  \left(\frac{m-2|x|}{m}\right)^2
\leq\frac{w^2p}{m^2},
\end{equation}
since $|m-2|x||\leq w$ throughout the band.For $w\leq\sqrt m$, \cref{eq:band-width-estimates}
tells us $w^2=O(mp^2)$, and hence, we have
$p\|\bar z\|^2=O(p^3/m)$. For $w>\sqrt m$, we get
\[
p\|\bar z\|^2
\leq
\frac1{2^m}\sum_{x\in\{0,1\}^m}
\left(\frac{m-2|x|}{m}\right)^2
=\frac1m.
\]
This is because for all $x$, $|x|$ has mean $m/2$ and variance $m/4$. Since $p=\Theta(1)$ in this range, we conclude that
\begin{equation}
\label{eq:band-moment-bound}
p\|\bar z\|^2=O\!\left(\frac{p^3}{m}\right)
\end{equation}
for all $w$. 

\paragraph{The larger-$\varepsilon$ regime.}

We first consider the range $n^{-1/2}\leq\varepsilon\leq1/128$. Let $m$ be the largest odd integer less than $n/2$, and let
$d=n-m$. Then $m=\Theta(n)$ and $d\geq n/2$.
By \cref{eq:bound_on_beta_and_pmax}, for sufficiently large $n$, we can find blocks $U_1,\ldots,U_R$ satisfying
\[
\frac1{64}\leq\beta\leq1,
\qquad
p_{\max}\leq2^{-\lfloor\sqrt d/10\rfloor}.
\]
These are the blocks used to define the distributions  $\mathcal D_{\textit{yes}}$ and $\mathcal D_{\textit{no}}$.

Next, let $\alpha=\rho=1$. Under $\mathcal D_{\textit{no}}$, every block is then assigned a function $f_i^-$. By \cref{eq:global-distance-from-monotonicity}, every sampled function therefore has distance $\beta\delta_w$ from monotonicity. Next, choose $w$ to be the smallest odd integer in $\{1,\ldots,m\}$ such that
\[
\beta\delta_w\geq\varepsilon.
\]
Such a choice must exist because $\delta_m=1/2$ and
\[
\beta\delta_m=\frac{\beta}{2}
\geq\frac1{128}\geq\varepsilon.
\]
Furthermore, every function sampled from $\mathcal D_{\textit{no}}$ is $\varepsilon$-far from monotone. 
    
Next, we bound the mass of the chosen band.  If $w>1$, the minimality of $w$ tells us that
\[
\delta_{w-2}<\frac{\varepsilon}{\beta}.
\]
Increasing the width from $w-2$ to $w$ adds the following terms to the sum defining $\delta_w$:
\[
\begin{aligned}
\delta_w-\delta_{w-2}
&=\frac1{2^m}
  \rev{\Bigl(}\binom{m-1}{L}+\binom{m-1}{U-1}\rev{\Bigr)} \leq O(m^{-1/2}).
\end{aligned}
\]
Therefore, 
\[
\delta_w
=O\!\left(\frac{\varepsilon}{\beta}+\frac1{\sqrt m}\right).
\]
The same bound holds when $w=1$ since $\delta_1=O(m^{-1/2})$. Since $m = \Theta(n)$ and by assumption, $\varepsilon\geq n^{-1/2}$, we have $m^{-1/2}=O(\varepsilon)$. Using $\beta\leq1$ and
\cref{eq:band-mass-distance-comparison,eq:band-moment-bound},
we get
\[
p\leq4\delta_w=O(\varepsilon/\beta),
\qquad
p\|\bar z\|^2
=O\!\left(\frac{\varepsilon^3}{\beta^3m}\right).
\]

Finally, substituting these estimates into
\cref{eq:general-ensemble-distance} gives
\begin{equation}
\begin{aligned}
\bigl\|\rho_{\textit{yes}}^{(t)}
      -\rho_{\textit{no}}^{(t)}\bigr\|_1^2
&\leq
64t\beta p\|\bar z\|^2
+32t^2p^2\beta p_{\max}\\
&=
O\!\left(
\frac{t\varepsilon^3}{\beta^2m}
+\frac{t^2\varepsilon^2p_{\max}}{\beta}
\right)\\
&=
O\!\left(
\frac{t\varepsilon^3}{n}
+t^2\varepsilon^2p_{\max}
\right),
\end{aligned}
\end{equation}
where we have used that $\alpha=\rho=1$ as well as 
$\beta\geq1/64$ and $m=\Theta(n)$. The second term is negligible since $p_{\max}\leq2^{-\lfloor\sqrt d/10\rfloor}$ and $d\geq n/2$. Therefore, in order for the trace norm to be a constant, we need $t = \Omega(n/\varepsilon^3)$, yielding our lower bound. 

\paragraph{The smaller-$\varepsilon$ regime.}

We now consider the range
$2^{-\sqrt n/200}\leq\varepsilon<n^{-1/2}$. Take $w = 1$ and choose an odd $m=\Theta(n)$ such that $m \leq n/2$. Furthermore, let $d = n-m$, and as before, fix blocks satisfying \cref{eq:bound_on_beta_and_pmax} so that 
\[
\frac1{64}\leq\beta\leq1,
\qquad
p_{\max}\leq2^{-\lfloor\sqrt d/10\rfloor}.
\]
We choose $m$ sufficiently small so that
\[
\beta\delta_1\geq\frac{12}{\sqrt n}.
\]
Such a choice is possible since $\delta_1=\Theta(m^{-1/2})$ and the lower bound
$\beta\geq1/64$ holds for the chosen blocks.

Next, let 
\[
\alpha=\frac{1}{3},
\qquad
\rho=\frac{2\varepsilon}{\beta\delta_1}.
\]
Since $\beta\delta_1\geq12/\sqrt n$ and
$\varepsilon<n^{-1/2}$, it follows that
\[
0<\rho\leq\frac{\varepsilon\sqrt n}{6}<\frac16<\alpha,
\]
so these parameters define valid distributions.
By \cref{eq:global-distance-from-monotonicity,eq:expected-negative-mass}, the expected distance of a function sampled from $\mathcal D_{\textit{no}}$ is
\[
\beta\rho\delta_1=2\varepsilon.
\]
From  \cref{eq:no-farness-chebyshev}, we have
\[
\Pr_{f\sim\mathcal D_{\textit{no}}}\!\left[
\operatorname{dist}(f,\rev{\mathrm{MON}}_{d+m})<\varepsilon
\right]
\leq
\frac{4p_{\max}}{\rho\beta}
=
\frac{2\delta_1p_{\max}}{\varepsilon}
=o(1).
\]
The last step holds in this regime, since
$\delta_1\leq1/2$, $p_{\max}\leq2^{-\lfloor\sqrt d/10\rfloor}$, $d\geq n/2$, and $\varepsilon\geq2^{-\sqrt n/200}$. Therefore, a function sampled from $\mathcal D_{\textit{no}}$ is $\varepsilon$-far from monotone with probability $1-o(1)$. 

For $w=1$, \cref{eq:band-width-estimates,eq:band-moment-bound}
give
\[
p,\delta_1=\Theta(m^{-1/2}),
\qquad
p\|\bar z\|^2=O(m^{-5/2}).
\]
Since $\alpha=1/3$,
\[
\frac{\rho^2}{1-\alpha+\rho}
=\frac{\rho^2}{2/3+\rho}
\leq\frac32\rho^2.
\]
Applying \cref{eq:general-ensemble-distance} and substituting
$\rho=2\varepsilon/(\beta\delta_1)$, we obtain
\begin{equation}
\begin{aligned}
\bigl\|\rho_{\textit{yes}}^{(t)}
      -\rho_{\textit{no}}^{(t)}\bigr\|_1^2
&\leq
96\rho^2t\beta p\|\bar z\|^2
+32t^2p^2\beta p_{\max}\\
&=
\frac{384t\varepsilon^2p\|\bar z\|^2}{\beta\delta_1^2}
+32t^2p^2\beta p_{\max}\\
&=
O\!\left(
\frac{t\varepsilon^2}{m^{3/2}}
+\frac{t^2p_{\max}}{m}
\right)\\
&=O\!\left(
\frac{t\varepsilon^2}{n^{3/2}}
+\frac{t^2p_{\max}}{n}\right),
\end{aligned}
\end{equation}
where we have used that $\beta=\Theta(1)$ and $m=\Theta(n)$. The second term is negligible and hence for a constant trace norm, we require $t=\Omega(n^{3/2}/\varepsilon^2)$.

Combining the two regimes, we conclude that for
$2^{-\sqrt n/200}\leq\varepsilon\leq1/128$,
monotonicity testing with two-sided error at most $1/3$ requires
\begin{equation}
\label{eq:final-copy-lower-bound}
t=\Omega\!\left(
\min\left\{
\frac{n}{\varepsilon^3},
\frac{n^{3/2}}{\varepsilon^2}
\right\}
\right)
\end{equation}
copies of the uniform function state. 

\section{Lower bounds for additional Boolean properties}
\label{sec:lower-bounds-for-additional-properties}

We adapt the construction from \Cref{sec:two-sided-lower-bounds} to prove lower bounds for testing union-closedness and $2$-monotonicity. We use the same selector blocks but with different local functions inside them.  A Boolean function $f:\{0,1\}^n\to\{0,1\}$ is \emph{union-closed} if, for all $x,y\in\{0,1\}^n$,
\[
    f(x)=f(y)=1
    \quad\Longrightarrow\quad
    f(x\vee y)=1,
\]
where $(x\vee y)_i:=\max\{x_i,y_i\}$ for every $i\in[n]$. The function is \emph{$2$-monotone} if there are no
vertices $x\prec y\prec z$ such that
\[
    (f(x),f(y),f(z))=(1,0,1)
\]
where $x\prec y$ means that $x\neq y$ and
$x_i\leq y_i$ for every $i\in[n]$. 

\begin{proof}[Proof of \Cref{thm:intro-other}]
Let $d:=\lfloor n/2\rfloor$ and $m:=n-d$. Write
each input as $(y,x)\in\{0,1\}^d\times\{0,1\}^m$.
Next, we fix the selector blocks $U_1,\ldots,U_R$ from
\Cref{subsec:block-construction}. Recall that $p_j=|U_j|/2^d$ is the fraction of selector
inputs belonging to $U_j$. By
\Cref{eq:bound_on_beta_and_pmax}, we have
\[
    \beta=\sum_j p_j\geq\frac{1}{64},
    \qquad
    p_{\max}=\max_j p_j\leq2^{-r},
\]
where $r=\lfloor\sqrt d/10\rfloor$. We will use the same fixed blocks for both properties.

For each pair of distinct coordinates $a,b\in[m]$
of the local input $x\in\{0,1\}^m$, define
\[
    D_{ab}(x):=x_a\oplus x_b,
    \qquad
    E_{ab}(x):=1-D_{ab}(x).
\]
Thus, $D_{ab}(x)=1$ when $x_a$ and $x_b$ differ,
whereas $E_{ab}(x)=1$ when they agree.
Independently for each block $U_j$, choose a uniformly
random unordered pair $\{a_j,b_j\}\subseteq[m]$. For each property, the assignments below define
distributions $\mathcal D_{\textit{yes}}$ and
$\mathcal D_{\textit{no}}$ over global functions. We write $f_{\textit{yes}}\sim\mathcal D_{\textit{yes}}$
and $f_{\textit{no}}\sim\mathcal D_{\textit{no}}$
for the sampled functions. The coordinate pairs are chosen once when selecting the function and remain fixed across all copies.

For union-closedness, set
\[
    f_{\textit{yes}}(y,x):=E_{a_jb_j}(x),
    \qquad
    f_{\textit{no}}(y,x):=D_{a_jb_j}(x)
    \qquad\text{when }y\in U_j.
\]
Both functions equal $0$ when $y$ satisfies no selector
term and equal $1$ when $y$ satisfies at least two
selector terms. The independent random choices of
$\{a_j,b_j\}$ define the distributions
$\mathcal D_{\textit{yes}}$ and $\mathcal D_{\textit{no}}$.

We will show that every function sampled from
$\mathcal D_{\textit{yes}}$ is union-closed.
Suppose $f_{\textit{yes}}(y,x)=f_{\textit{yes}}(y',x')=1$. 
This implies that both $y$ and $y'$ satisfy at least one selector term, and $y\vee y'$ satisfies every term satisfied by either. If $y\vee y'$ satisfies at least two terms, then $f_{\textit{yes}}(y\vee y',x\vee x')=1$ by definition. Otherwise, $y$, $y'$, and $y\vee y'$ belong to the same block $U_j$. Since
$E_{a_jb_j}(x)=E_{a_jb_j}(x')=1$, we have
\[
    (x\vee x')_{a_j}
    =x_{a_j}\vee x'_{a_j}
    =x_{b_j}\vee x'_{b_j}
    =(x\vee x')_{b_j}.
\]
Therefore, $E_{a_jb_j}(x\vee x')=1$ and
$f_{\textit{yes}}(y\vee y',x\vee x')=1$
in this case as well.

Next, we show that every function sampled from
$\mathcal D_{\textit{no}}$ is far from union-closed.
Fix $y\in U_j$ and all coordinates of $x$ except
$a_j,b_j$. Setting $(x_{a_j},x_{b_j})$ to
$(1,0)$, $(0,1)$, and $(1,1)$ gives three inputs
with labels $1,1,0$, respectively. The union of the
first two inputs is the third, so any union-closed
function must differ from $f_{\textit{no}}$ on at
least one of these inputs. For each fixed $y$, varying the other $m-2$ coordinates gives $2^{m-2}$ disjoint violating triples. Furthermore, triples from different $U_j$ are also disjoint. Therefore, making $f_{\textit{no}}$ union-closed requires changing at least 
\[
    \frac{2^{m-2}\sum_j|U_j|}{2^{d+m}}
    =\frac{\beta}{4}
    \geq\frac{1}{256}
\]
of its values. In particular, every function sampled
from $\mathcal D_{\textit{no}}$ is $\varepsilon$-far
from union-closed for the stated range of $\varepsilon$.

We now bound the trace distance between the averaged
$t$-copy ensembles. Let $\rho_b^{(t)}$ denote the averaged
state of $t$ copies under $\mathcal D_b$, for
$b\in\{\textit{yes},\textit{no}\}$.
By \eqref{eq:rho_yes_and_no_t_copies}, we have
\[
    \bigl\|\rho_{\textit{yes}}^{(t)}
          -\rho_{\textit{no}}^{(t)}\bigr\|_1^2
    \leq
    32\sum_{j=1}^R\bigl(1-F_{2tp_j}\bigr),
\]
where
\[
    F_\mu:=
    F\bigl(\sigma_{\textit{yes}}(\mu),
           \sigma_{\textit{no}}(\mu)\bigr)
\]
is the fidelity between the local averaged states.
It remains to bound $1-F_\mu$. Let $K:=\binom{m}{2}$ be the number of possible coordinate pairs. Since each pair is chosen uniformly at random, the local averaged states are
\[
\begin{aligned}
    \sigma_{\textit{yes}}(\mu)
    &=
    \frac{1}{K}\sum_{1\leq a<b\leq m}
    \ket{\Phi_\mu(E_{ab})}\bra{\Phi_\mu(E_{ab})},\\
    \sigma_{\textit{no}}(\mu)
    &=
    \frac{1}{K}\sum_{1\leq a<b\leq m}
    \ket{\Phi_\mu(D_{ab})}\bra{\Phi_\mu(D_{ab})}.
\end{aligned}
\]
We can calculate $F_\mu$ using the overlaps between
these component vectors.

For any two coordinate pairs $\{a,b\}$ and $\{a',b'\}$,
we have
\[
    \operatorname{dist}(E_{ab},D_{a'b'})
    =
    \begin{cases}
        1, & \{a,b\}=\{a',b'\},\\
        \frac{1}{2}, & \{a,b\}\neq\{a',b'\}.
    \end{cases}
\]
If the pairs are identical, the two functions are the complement of each other. Otherwise, there is at least a single coordinate that belongs to one of the pairs, and flipping this coordinate changes only one of the two function values. Therefore, the distance is $\frac{1}{2}$. 
It follows that
\[
    \langle\Phi_\mu(E_{ab})\mid\Phi_\mu(D_{a'b'})\rangle
    =
    \begin{cases}
        e^{-\mu}, & \{a,b\}=\{a',b'\},\\
        e^{-\mu/2}, & \{a,b\}\neq\{a',b'\}.
    \end{cases}
\]

The weighted overlap matrix of the ensemble states is
\[
    G_\mu
    :=
    \frac{1}{K}
    \left[
        \bigl(e^{-\mu}-e^{-\mu/2}\bigr)I_K
        +e^{-\mu/2}J_K
    \right],
\]
where $I_K$ is the identity matrix and $J_K$ is the
matrix whose entries are all $1$. Applying \Cref{eq:mixture-fidelity-overlap} to the local
averaged states gives
\[
    F_\mu=\|G_\mu\|_1.
\]
The eigenvalues of $G_\mu$ are
\[
    \frac{e^{-\mu}+(K-1)e^{-\mu/2}}{K},
    \qquad
    \frac{e^{-\mu}-e^{-\mu/2}}{K},
\]
with multiplicities $1$ and $K-1$ respectively. Hence,
\[
\begin{aligned}
    F_\mu
    &=
    \frac{e^{-\mu}+(K-1)e^{-\mu/2}}{K}
    +(K-1)\frac{e^{-\mu/2}-e^{-\mu}}{K}\\
    &=
    \frac{2(K-1)e^{-\mu/2}-(K-2)e^{-\mu}}{K}.
\end{aligned}
\]
Rearranging the expression for $F_\mu$, we get
\begin{equation}
\label{eq:union-local-fidelity-deficit}
\begin{aligned}
    1-F_\mu
    &=
    \bigl(1-e^{-\mu/2}\bigr)^2
    +\frac{2}{K}e^{-\mu/2}
        \bigl(1-e^{-\mu/2}\bigr)\\
    &\leq
    \frac{\mu^2}{4}+\frac{\mu}{K},
\end{aligned}
\end{equation}
where we have used that $e^{-\mu/2}\leq1$ and
$0\leq1-e^{-\mu/2}\leq\frac{\mu}{2}$. Applying \eqref{eq:rho_yes_and_no_t_copies} and substituting
$\mu=2tp_j$ into \Cref{eq:union-local-fidelity-deficit}
gives
\begin{equation}
\label{eq:union-global-distance}
\begin{aligned}
    \bigl\|\rho_{\textit{yes}}^{(t)}
          -\rho_{\textit{no}}^{(t)}\bigr\|_1^2
    &\leq
    32\sum_{j=1}^R
    \left(t^2p_j^2+\frac{2tp_j}{K}\right)\\
    &=
    32t^2\sum_{j=1}^R p_j^2
    +\frac{64t}{K}\sum_{j=1}^R p_j\\
    &\leq
    32\beta p_{\max}t^2+\frac{64\beta t}{K}.
\end{aligned}
\end{equation}
where we used that $\sum_jp_j=\beta$ and
\[
    \sum_jp_j^2
    \leq p_{\max}\sum_jp_j
    =\beta p_{\max}.
\]
Using $\beta\leq1$, $p_{\max}\leq2^{-\Omega(\sqrt n)}$,
and $K=\Theta(n^2)$, we have
\[
    32\beta p_{\max}t^2+\frac{64\beta t}{K}
    \leq
    32t^2 2^{-\Omega(\sqrt n)}
    +O\!\left(\frac{t}{n^2}\right).
\]
Therefore, any algorithm that distinguishes the two cases with probability $\frac{1}{3}$ requires at least $t=\Omega(n^2)$ copies, proving the lower bound. 

For $2$-monotonicity, the construction is identical except we define
\[
    f_{\textit{yes}}(y,x)=D_{a_jb_j}(x),
    \qquad
    f_{\textit{no}}(y,x)=E_{a_jb_j}(x)
    \qquad(y\in U_j),
\]
and we set both functions to $0$ whenever
$y\notin\bigcup_j U_j$. 

To show that $f_{\textit{yes}}$ is $2$-monotone, consider any chain such that
\[
    (y_0,x_0)\prec(y_1,x_1)\prec(y_2,x_2)
\]
where the first and last vertices have function value $1$. We will show that the value of $(y_1, x_1)$ must also be $1$. The selector inputs $y_0$ and $y_2$ lie in the same
block $U_j$, since the function is $0$ outside the
blocks and different blocks are incomparable. Therefore, $y_1$ has to belong to the same block. Within $U_j$, the function $f_{\mathrm{yes}}$ is the XOR of coordinates $a_j$ and $b_j$. Since $f_{\mathrm{yes}}(y_0,x_0)=1$, the pair
\[
    (x_{0,a_j},x_{0,b_j})
\]
is either $(0,1)$ or $(1,0)$. If it is $(0,1)$, the corresponding pair for $x_2$ is either $(0,1)$ or $(1,1)$. The latter is impossible since $f_{\mathrm{yes}}(y_2,x_2)=1$, so $x_2$ must also have selected bits $(0,1)$. Since $x_0\preceq x_1\preceq x_2$, we have that the corresponding bits of $x_1$ are necessarily $(0,1)$ and $f_{\mathrm{yes}}(y_1,x_1)=1$. The case of $(1,0)$ is analogous. Therefore, there cannot exist a chain that violates $2$-monotonicity.  

We can now show that every function sampled from
$\mathcal D_{\textit{no}}$ is $\varepsilon$-far from
$2$-monotone. Fix $y\in U_j$ and all coordinates of $x$ except $a_j,b_j$. Letting the selected pair be
$(0,0)$, $(1,0)$, and $(1,1)$ gives a chain on which
$f_{\textit{no}}$ has labels $(1,0,1)$. Therefore, any $2$-monotone function must differ from $f_{\textit{no}}$ on at least one vertex of this chain. Since the triples are disjoint as we vary the remaining coordinates, we can reuse the same counting argument as for union-closedness. Therefore, making $f_{\textit{no}}$ a $2$-monotone function requires changing at least $\beta/4\geq1/256$ fraction of its values, proving the claim. 

Finally, since the local averaged states are identical to those used for union-closedness (but with the ensembles exchanged), the fidelity is unchanged. Therefore, we get a $\Omega(n^2)$ copy lower bound for testing $2$-monotonicity.

\end{proof}

\paragraph{AI Disclosure} The author used GPT-6 Astra and GPT-5.1 in a role similar to that of a research collaborator. The author developed the initial ideas and arguments for the problem, which were subsequently strengthened and simplified after discussions with the models. The author takes full responsibility for the results and manuscript.

\printbibliography

\clearpage
\appendix
\section{Supporting proofs for the upper bounds}
\label{app:upper-bound-proofs}
This appendix gives the detailed proofs of the path-soundness and
path-variance lemmas used
in \Cref{sec:upper-bounds}.

\subsection{Proof of path soundness}
\label{app:path-soundness}
We prove \Cref{lem:path-soundness}, with the band, degree scales,
and path experiment defined in \Cref{sec:path-upper}.
\begin{proof}

Let $\tilde f$ agree with $f$ on $\mathcal B$, equal $0$ below the band, and equal $1$ above the band. Every violation
of $\tilde f$ lies entirely within $\mathcal B$ and is also
a violation of $f$. Since the procedure counts only violations within $\mathcal B$ and the two functions agree on $\mathcal B$, we have 
\[
    p_t(f)=p_t(\tilde f)
    \qquad\text{for every }t\geq1.
\]
We will work with $\tilde f$ for the remainder of the proof. Note that Hoeffding's inequality gives
\[
    \Pr_{x\sim\mathrm{Unif}(\{0,1\}^n)}
        [x\notin\mathcal B]
    \leq 2\exp\left(-\frac{2w^2}{n}\right)
    \leq \frac{\varepsilon}{50}.
\]
Therefore, $\tilde f$ differs from $f$ on at most $\varepsilon/50$ fraction of the inputs. Using the triangle inequality, we have
\[
    \dist(\tilde f,\mathrm{MON}_n)
    \geq \dist(f,\mathrm{MON}_n)-\frac{\varepsilon}{50}
    \geq \frac{49\varepsilon}{50}
    \geq \frac{\varepsilon}{2}.
\]

We also prove that $\tilde f$ has low-influence if $f$ has low-influence. An edge on which $\tilde f$ changes but $f$ does not must cross one of the two boundaries of $\mathcal B$. The total number of edges crossing the two boundaries is at most 
\[
2n\max_{0\leq k\leq n}\binom nk
    \leq 2^{n+2}\sqrt n
\]
Furthermore, since each undirected edge on which a Boolean function changes contributes $2/2^n$ to the influence (since the edge can be counted at either endpoint), the two boundaries increase the total influence by at most $8 \sqrt{n}$. We get
\begin{equation}
\label{eq:tilde_f_bound}
\mathbf I[\tilde f]
\leq \mathbf I[f]+8\sqrt n
\leq 10\sqrt n.  
\end{equation}

Next, let us apply \cref{lem:degree-selection} to $\tilde f$. For some $d \in \mathcal D_n$, we can find a subgraph 
$G=(A,B,E)$ satisfying
\[
    \deg_G(y)=d\quad(y\in B),
    \qquad
    \deg_G(x)\le2d\quad(x\in A),
    \qquad
    \sigma:=\frac{|B|}{2^n}
    \ge c\varepsilon d^{-5/8},
\]
where $c>0$ and $E \subseteq E_{\tilde f}^{-}$ (see \Cref{eq:set_of_violated_edges}). Since all edges in $E_{\tilde f}^{-}$ must have both endpoints in $\mathcal B$, all vertices of $G$ also lie in $\mathcal B$. 

We follow the analysis in \cite{DBLP:journals/siamcomp/KhotMS18}, adapting their argument to our modified band and walk lengths. We first consider the case $t_d=1$. Note, there are $2^n$ possible starting vertices, two possible directions, and at most $n$ coordinates per direction. Furthermore, there are $d |B|$ violated edges in the subgraph which can be selected in either direction. Hence, we have 
\[
    p_1(\tilde f)
    \geq \frac{2d|B|}{2^n\cdot 2n}
    =\frac{d|B|}{2^n n}
    \geq c \frac{\varepsilon d^{3/8}}n,
\]
proving the claim. 

Next, assume that $t_d \geq 2$, and let $c_0 \leq 1$. By definition, we have
\[
t_d =\left\lfloor c_0\varepsilon\sqrt n\,d^{-5/8}\right\rfloor 
\le c_0\varepsilon\sqrt n\,d^{-5/8} 
\le\varepsilon\sqrt n.
\]

We will show that 
every vertex reachable by at most $t_d$ steps from $\mathcal B$ has Hamming weight in $[n/4,3n/4]$. Since $2\le t_d\le\varepsilon\sqrt n$, we have
\begin{equation}
\label{eq:epsilon-log-bounds}
    \varepsilon\ge n^{-1/2},
    \qquad
    \log\frac{100}{\varepsilon}\le\log(100\sqrt n).
\end{equation}
Using the definition of $w$ and that $\varepsilon \leq 1/2$, we have
\begin{equation}
\label{eq:bound_on_w+t_d}
    \frac{w+t_d}{n}
    \le
    \sqrt{\frac{\log(100\sqrt n)}{2n}}
    +\frac1n+\frac1{2\sqrt n}
    \longrightarrow 0
    \qquad\text{as }n\to\infty.
\end{equation}

Therefore, choosing $n_0$ sufficiently large, we have $w+t_d\le n/4$ for all $n\ge n_0$. For $y \in \mathcal{B}$, let $u$ be obtained from $y$ by at most $j$ upward or downward steps where $0 \leq j \leq t_d$. Then, we have the following bound on the Hamming weight of $u$:
\[
    \frac{n}{4}
    \le \frac{n}{2}-(w+t_d)
    \le |u|
    \le \frac{n}{2}+(w+t_d)
    \le \frac{3n}{4}.
\]
For the remainder of the proof, we will consider the case where $B$ consists of upper endpoints labeled $0$ and $A$ consists of lower endpoints labeled $1$. The opposite case will follow by symmetry (using downward walks rather than upward walks). 

We would like to upper-bound the probability
\[
    \Pr\!\left[
        Y\in\mathcal B,\
        \tilde f(Y)\ne
        \tilde f(Y^{\uparrow(t_d-1)})
    \right],
\]
where $Y$ is sampled uniformly at random from $\{0,1\}^n$, and for $Y\in\mathcal B$, $Y^{\uparrow(t_d-1)}$ is obtained by flipping a random set of $t_d-1$ zero coordinates of $Y$ to ones. Given $Y \in \mathcal B$, denote the upward walk as
\[
    U_0=Y,\ U_1,\ldots,U_{t_d-1}
    =Y^{\uparrow(t_d-1)},
\]
where each $U_{j+1}$ is obtained from $U_j$ by changing a randomly chosen zero coordinate to one. We will first bound the probability of reaching a fixed vertex at an intermediate step of this walk. 

Consider $0 \leq j \leq t_{d} - 1$ and consider some vertex $u$ such that $|u| = k$ (where $k$ must be between $n/4$ and $3n/4$ by \eqref{eq:bound_on_w+t_d}.)  There are $\binom{k}{j}$ possible starting vertices, obtained by changing $j$ of the one coordinates of $u$ to zeros. Furthermore, each has $n-k+j$ zero coordinates where exactly one of the $\binom{n-k+j}{j}$ sets of $j$ coordinates leads to $u$. Since $Y$ is chosen uniformly at random, we have the following bound
\[
    \Pr[Y\in\mathcal B,\ U_j=u]
    \le
    \frac1{2^n}
    \frac{\binom{k}{j}}{\binom{n-k+j}{j}}.
\]
Next, observe that \[
\begin{aligned}
    \frac{\binom{k}{j}}{\binom{n-k+j}{j}}
    &=
    \prod_{h=0}^{j-1}\frac{k-h}{n-k+j-h}
    \le
    \left(\frac{k}{n-k+1}\right)^j\\
    &\le
    \left(\frac{n/2+w+t_d}{n/2-w-t_d}\right)^j\\
    &=
    \left(1+\frac{2(w+t_d)}{n/2-w-t_d}\right)^j\\
    &\le
    \left(1+\frac{8(w+t_d)}n\right)^j
    \le
    \exp\!\left(\frac{8t_d(w+t_d)}n\right),
\end{aligned}
\]
where we have used that $k\le n/2+w+t_d$, $w+t_d\le n/4$, and $j \leq t_d$. Using $t_d\leq\varepsilon\sqrt n$ and the definition of $w$,
\begin{equation}
\label{eq:bound_t_d_w_t_d}
\begin{aligned}
    \frac{t_d(w+t_d)}n
    &\leq \frac{\varepsilon}{\sqrt n}
    \left(\sqrt{\frac n2\log\frac{100}{\varepsilon}}
    +1+\varepsilon\sqrt n\right)\\
    &=\varepsilon\sqrt{\frac12\log\frac{100}{\varepsilon}}
    +\frac{\varepsilon}{\sqrt n}+\varepsilon^2
    =O(1),
\end{aligned}
\end{equation}
uniformly for $0<\varepsilon\leq1/2$. Therefore, we have
\[
    \Pr[Y\in\mathcal B,\ U_j=u]
    \le
    \frac1{2^n}
    \exp\!\left(\frac{8t_d(w+t_d)}n\right)
    \le \frac{C}{2^n}.
\]
for an absolute constant $C>0$, chosen uniformly for $c_0\leq1$.
We fix this constant before choosing $c_0$ below.

Next, let us bound the probability that a particular edge is chosen on an upward walk. Fix $0\le j\le t_d-2$ and consider an edge $\{u,v\}$ with $v = u+e_i$ for some $i$ with $u_i = 0$. If $u$ can be reached from $\mathcal B$ after $j$ steps then we have that $n - |u| \geq n/4$. Furthermore, if we can reach $u$, then the coordinate $i$ is chosen with probability $1/(n - |u|)$. Therefore, we have 
\[
\begin{aligned}
    \Pr[Y\in\mathcal B,\ U_j=u,\ U_{j+1}=v]
    &=
    \Pr[Y\in\mathcal B,\ U_j=u]\,
    \frac1{n-|u|}\\
    &\le
    \frac{C}{2^n}\cdot\frac4n
    =\frac{4C}{n2^n}.
\end{aligned}
\]
Note that if $u$ is not reachable, then this probability is zero and therefore, the bound holds. There are $2^{n-1}\mathbf I[\tilde f]$ edges on which $\tilde f$ changes. A union bound over these edges gives us for each $0\le j\le t_d-2$,
\[
\begin{aligned}
    \Pr[Y\in\mathcal B,\ \tilde f(U_j)\ne\tilde f(U_{j+1})]
    &\le
    2^{n-1}\mathbf I[\tilde f]\cdot\frac{4C}{n2^n}\\
    &=\frac{2C\mathbf I[\tilde f]}n
    \le\frac{20C}{\sqrt n},
\end{aligned}
\]
where we have used the influence bound
from \Cref{eq:tilde_f_bound}. If $\tilde f(Y)\ne\tilde f(U_{t_d-1})$, then the function value must have changed during at least a single step. Taking a union bound over the $t_d-1$ steps gives us
\[
\begin{aligned}
    \Pr[Y\in\mathcal B,\ \tilde f(Y)\ne \tilde f(Y^{\uparrow(t_d-1)})]
    &\le
    \sum_{j=0}^{t_d-2}
    \Pr[Y\in\mathcal B,\
        \tilde f(U_j)\ne\tilde f(U_{j+1})]\\
    &\le
    \frac{20C(t_d-1)}{\sqrt n}
    \le \frac{20Ct_d}{\sqrt n}.
\end{aligned}
\]

Next, call a vertex $y \in B$ persistent if
\[
    \Pr[\tilde f(y^{\uparrow(t_d-1)})=0]\ge\frac9{10}.
\]
Let $B_{\mathrm{bad}}$ be the set of nonpersistent vertices in $B$. Since $\tilde{f}(y) = 0$ for every $y \in B$, each vertex in $B_{\mathrm{bad}}$ has a probability greater than $1/10$ of a change in the function value. Since $Y$ is uniform and $B\subseteq\mathcal B$, we have
\[
    \frac{|B_{\mathrm{bad}}|}{10\cdot2^n}
    \le
    \Pr[Y\in\mathcal B,\
        \tilde f(Y)\ne\tilde f(Y^{\uparrow(t_d-1)})]
    \le \frac{20Ct_d}{\sqrt n},
\]
which implies 
\[
    \frac{|B_{\mathrm{bad}}|}{2^n}
    \le \frac{200Ct_d}{\sqrt n}.
\]
Using $t_d\le c_0\varepsilon\sqrt n\,d^{-5/8}$ and
$\sigma\ge c\varepsilon d^{-5/8}$, we obtain
\[
    \frac{|B_{\mathrm{bad}}|}{2^n}
    \le 200Cc_0\varepsilon d^{-5/8}
    \le \frac{200Cc_0}{c}\,\sigma
    \le \frac{\sigma}{100},
\]
where the last inequality holds provided we choose
$c_0\le c/(20000C)$. This implies that at least $99/100 |B|$ vertices are persistent. Let $B':=B\setminus B_{\mathrm{bad}}$ and let $E' \subseteq E$ consist of those edges incident to $B'$. Since every vertex in $B$ has degree $d$, we note that
\[
    |E'|=d|B'|\ge\frac{99}{100}d|B|.
\]
Therefore, every edge in $E'$ has a persistent vertex in $B'$, and every vertex in $A$ is incident to at most $2d$ edges in $E'$. 

We can now lower bound $p_{t_d}(\tilde f)$, adapting
\cite[Section~7.3]{DBLP:journals/siamcomp/KhotMS18}. Fix $x\in A$, and let $m_x \leq 2d$ be its degree in the subgraph $(A,B',E')$. Next, let $r:= n - |x|$. By definition of $\mathcal B$ and \Cref{eq:bound_on_w+t_d}, we have that
\[
    r\ge \frac n4\ge t_d.
\]
Next, let $S$ be a uniformly random $t_d$ element subset of the zero coordinates of $x$ and $y$ be the corresponding vertex obtained by flipping these coordinates from zero to one. 

For each neighbor of $x$, that is $v_i = x+e_i$, define 
\[
    H_i:=\{i\in S,\ \tilde f(y)=0\}.
\]
Note that $\Pr[i\in S]=t_d/r$. Furthermore, conditional on $i \in S$, the set $S\setminus\{i\}$ is a uniformly random
$(t_d-1)$-element subset of the zero coordinates of $v_i$. Thus $y$ has the conditional distribution of $v_i^{\uparrow(t_d-1)}$. Since we know that $v_i$ is persistent, we have 
\[
\begin{aligned}
    \Pr[H_i]
    &=\Pr[i\in S]\,
      \Pr[\tilde f(y)=0\mid i\in S]\\
    &\ge \frac{9t_d}{10r}.
\end{aligned}
\]
We would like to bound $\Pr[\tilde f(y)=0]$ which we know is bounded below by $\Pr\!\left[ \bigcup_i H_i\right]$. Note that since 
\[
    \Pr[H_i\cap H_j]
    \le \Pr[i,j\in S]
    =\frac{t_d(t_d-1)}{r(r-1)},
\]
we have 
\[
\begin{aligned}
    \Pr[\tilde f(y)=0]
    &\ge \Pr\!\left[\bigcup_i H_i\right]\\
    &\ge \sum_i\Pr[H_i]
          -\sum_{i<j}\Pr[H_i\cap H_j]\\
    &\ge \frac{9}{10}\frac{m_x t_d}{r} - \binom{m_x}{2} \frac{t_d(t_d-1)}{r(r-1)}\\
    &\ge \frac{9}{10}\frac{m_x t_d}{r} - (\frac{m_x t_d}{r})^2.
\end{aligned}
\]

Using $m_x\leq 2d$, $r\geq n/4$, and
$t_d\leq c_0\varepsilon\sqrt{n}\,d^{-5/8}$, we obtain
\[
    \frac{m_x t_d}{r}
    \leq \frac{8c_0\varepsilon d^{3/8}}{\sqrt{n}}
    \leq 4c_0 n^{-1/8}
    \leq 4c_0,
\]
where the second inequality uses $\varepsilon\leq 1/2$
and $d\leq n$. Therefore, provided we choose $c_0\leq 1/40$, we have the following lower bound $\Pr[\tilde f(y)=0]\geq 4m_x t_d/(5n)$. If $\tilde f(y)=0$ in this upward walk, then $y$ cannot lie above
the band, where $\tilde f$ equals $1$, and it cannot lie below the band because
$x\in\mathcal B$ and $y\succeq x$. Thus this event is counted by the path experiment.
Finally, to bound the total probability, since the procedure chooses each starting vertex with probability $2^{-n}$ and the upward walk with probability $1/2$, we have 

\[
\begin{aligned}
    p_{t_d}(\tilde{f})
    &\ge \frac1{2^{n+1}}
          \sum_{x\in A}\frac{4m_xt_d}{5n}\\
    &=\frac{2t_d|E'|}{5n2^n}\\
    &\ge \frac{99}{250}\frac{\sigma t_d d}{n}\\
    &\ge \frac{99c}{250}
          \frac{\varepsilon t_d d^{3/8}}{n}.
\end{aligned}
\]
In case the graph $G = (A, B, E)$ is oriented in the reverse direction where $B$ are the lower endpoints and $A$ are the upper endpoints, the same argument holds but with downward walks. Choosing 
\[
    0<c_0\le
    \min\left\{\frac1{40},\frac{c}{20000C}\right\}
    \qquad
    c_1:=\frac c4,
\]
and using that $p_{t_d}(\tilde{f}) = p_{t_d}(f)$, we conclude that 
\[
    p_{t_d}(f)\ge
    \frac{c_1\varepsilon t_d d^{3/8}}{n}.
\]

\end{proof}

\subsection{Second moments of the path observables}
\label{app:path-variance}
\begin{proof}[Proof of \Cref{lem:path-estimator-variance}]
Since the two batches are independent,
\[
\begin{aligned}
    \operatorname{Var}(\widehat p_d)
    &=\frac{\operatorname{Var}(U_d)}{4m}
      +\frac{\operatorname{Var}(V_d)}{16m}\\
    &\leq\frac{\mathbb E[U_d^2]}{4m}
      +\frac{\mathbb E[V_d^2]}{16m}.
\end{aligned}
\]
It therefore suffices to bound each of the two second moments by $O(t_d/\sqrt n)$.

We first bound $\mathbb E[U_d^2]$. Let us define
\begin{equation}
\label{eq:lambda_d_equation}
    \Lambda_d:=
    \max_x\left(
        \sum_y\kappa_d(x,y)+\sum_y\kappa_d(y,x)
    \right).
\end{equation}
Since $\ket{\xi_{xy}}$ is a unit vector in the span of $\ket{x}\otimes\ket{-}$ and $\ket{y}\otimes\ket{-}$,
\[
    \ket{\xi_{xy}}\bra{\xi_{xy}}
    \preceq
    \bigl(\ket{x}\bra{x}+\ket{y}\bra{y}\bigr)
    \otimes\ket{-}\bra{-}.
\]
Multiplying by $\kappa_d(x,y)$ and summing over all $(x,y)\in\mathcal P_d$ gives us $0\preceq Q_d\preceq\Lambda_d I$. We may write
\begin{equation}
\label{eq:disagreement-second-moment}
    \mathbb E[U_d^2]
    \leq\Lambda_d\,\mathbb E[U_d]
    =\Lambda_d q_d(f),
\end{equation}
and it remains to bound $\Lambda_d$ and $q_d(f)$. 

To bound $\Lambda_d$, fix $x\in\mathcal B$ and write $k=|x|$. An upper endpoint of $x$ is obtained by choosing $t_d$ of the $n-k$ zero coordinates while a lower endpoint is obtained by choosing $t_d$ of the $k$ one coordinates. Therefore, there are  $\binom{n-k}{t_d}$ possible upper endpoints and $\binom{k}{t_d}$ possible lower endpoints. Using the definition of $\kappa_d$, we have
\begin{equation}
\label{eq:kappa_value_bound}
\begin{aligned}
    \sum_y\kappa_d(x,y)
    &\leq\frac12\left(
        1+\frac{\binom{n-k}{t_d}}{\binom{k+t_d}{t_d}}
    \right),\\
    \sum_y\kappa_d(y,x)
    &\leq\frac12\left(
        1+\frac{\binom{k}{t_d}}{\binom{n-k+t_d}{t_d}}
    \right).
\end{aligned}
\end{equation}
with equality only if the corresponding endpoints lie in $\mathcal B$. Since $n/2-w\leq k\leq n/2+w$ and $w\leq n/4$, it follows that 
\begin{equation}
\label{eq:binom_eq_bound}
\begin{aligned}
    \frac{\binom{n-k}{t_d}}{\binom{k+t_d}{t_d}}
    &=\prod_{j=0}^{t_d-1}\frac{n-k-j}{k+t_d-j}\\
    &\leq\left(\frac{n/2+w}{n/2-w}\right)^{t_d}\\
    &=\left(1+\frac{2w}{n/2-w}\right)^{t_d}\\
    &\leq\left(1+\frac{8w}{n}\right)^{t_d}\\
    &\leq\exp\!\left(\frac{8t_dw}{n}\right)
    =O(1).
\end{aligned}
\end{equation}
For $t_d\geq2$, the final bound follows from
\eqref{eq:bound_t_d_w_t_d}. For $t_d=1$, it follows directly from
$w=O(\sqrt{n\log n})$ in the present range $\varepsilon\geq n^{-1/2}$.
After increasing the absolute $n_0$ if necessary, $w+t_d\leq n/4$ in both cases. The second ratio satisfies the same bound by replacing $k$ with $n-k$. Both of the summations vanish outside
$\mathcal B$ so $\Lambda_d=O(1)$.

Next, we bound $q_d(f)$ using the estimate given in the proof of \cref{lem:path-soundness}. Let $X$ and $Y$ denote the starting
and final vertices sampled by the procedure. By the definition of $\kappa_d$, we have
\[
\begin{aligned}
    q_d(f)
    &=\Pr[X\in\mathcal B,\ Y\in\mathcal B,\ f(X)\ne f(Y)]\\
    &\leq\Pr[X\in\mathcal B,\ f(X)\ne f(Y)].
\end{aligned}
\]
We can reuse the argument from the proof of
\cref{lem:path-soundness}. The same binomial-ratio estimate applies to every $0\leq j\leq t_d$ in the present parameter range. For either the upward or downward direction, the probability of starting in $\mathcal B$ and traversing an edge on which $f$ changes at any specified step is at most $2C\mathbf I[f]/n$. Furthermore, taking $t_d$ steps instead of $t_d-1$
simply adds one term to the union bound. Therefore, we have 
\[
    q_d(f)
    \leq \frac{2Ct_d\mathbf I[f]}{n}
    \leq \frac{4Ct_d}{\sqrt n},
\]
where we used $\mathbf I[f]\leq2\sqrt n$. Combining this with $\Lambda_d=O(1)$ and
\Cref{eq:disagreement-second-moment}, we get
\[
    \mathbb E[U_d^2]=O(t_d/\sqrt n).
\]

Next, we bound $\mathbb E[V_d^2]$. Since $H_d\otimes Z$ is diagonal in the computational basis, we can sample $V_d$ by measuring one copy of $\ket{\psi_f}$ in this basis, and reporting the value $h_d(X)(-1)^{f(X)}$, where $X$ is uniformly random. Therefore, we may write
\[
    \mathbb E[V_d^2]
    =\mathbb E_X[h_d(X)^2]
    =\frac1{2^n}\sum_{x\in\mathcal B}h_d(x)^2,
\]
noting that $h_d(x)=0$ outside $\mathcal B$. It remains to bound the sum. 

We first consider vertices that are near the boundaries of the band -- in particular, the lowest and highest $t_d$ Hamming layers of $\mathcal B$. Note that each layer has probability at most $2/\sqrt{n}$. We then have
\[
    \frac{|\mathcal B_{\partial}|}{2^n}
    \leq\frac{4t_d}{\sqrt n}.
\]
By definition of $h_d$ and $\Lambda_d$ (see \Cref{eq:lambda_d_equation}),
\[
    |h_d(x)|
    \leq\sum_y\kappa_d(x,y)+\sum_y\kappa_d(y,x)
    \leq\Lambda_d 
\]
Using the earlier bound $\Lambda_d=O(1)$, the
contribution from the layers in $\mathcal B_{\partial}$ satisfies
\[
    \frac1{2^n}\sum_{x\in\mathcal B_{\partial}}h_d(x)^2
    \leq\Lambda_d^2\frac{|\mathcal B_{\partial}|}{2^n}
    \leq\frac{4\Lambda_d^2t_d}{\sqrt n}
    =O(t_d/\sqrt n).
\]

Consider any $x\in\mathcal B\setminus\mathcal B_{\partial}$, and let $k:=|x|$. Both layers of weights $k-t_d$ and $k+t_d$ lie in $\mathcal B$, and we can write
\[
\begin{aligned}
    h_d(x)
    &=\sum_y\kappa_d(x,y)-\sum_y\kappa_d(y,x)\\
    &=\frac{1}{2}\left(
        \frac{\binom{n-k}{t_d}}{\binom{k+t_d}{t_d}}
        -
        \frac{\binom{k}{t_d}}{\binom{n-k+t_d}{t_d}}
    \right).
\end{aligned}
\]
using the bounds in \Cref{eq:kappa_value_bound}. Note that for consecutive Hamming weights, we have
\[
\begin{aligned}
    \frac{\binom{n-\ell}{t_d}}{\binom{\ell+t_d}{t_d}}
    -\frac{\binom{n-\ell-1}{t_d}}{\binom{\ell+t_d+1}{t_d}}
    &=
    \frac{\binom{n-\ell}{t_d}}{\binom{\ell+t_d}{t_d}}
    \frac{t_d(n+1)}{(n-\ell)(\ell+t_d+1)}\\
    &\leq C\frac{t_d(n+1)}{(n/4)^2}\\
    &\leq 32C \frac{t_d}{n} ,
\end{aligned}
\]
where we have used the bound from \Cref{eq:binom_eq_bound} with the same uniform constant $C$ enlarged if necessary, as well as $n-\ell\geq n/4$, $\ell+t_d+1\geq n/4$ since $\ell$ lies in the band $\mathcal B$. For $k\leq n-k$, we can write
\[
\begin{aligned}
    2h_d(x)
    &=
    \sum_{\ell=k}^{n-k-1}
    \left(
        \frac{\binom{n-\ell}{t_d}}{\binom{\ell+t_d}{t_d}}
        -
        \frac{\binom{n-\ell-1}{t_d}}{\binom{\ell+t_d+1}{t_d}}
    \right)\\
    &\leq 32C (n-2k) \frac{t_d}{n}. 
\end{aligned}
\]
For $k>n-k$, we simply reverse the summation order to bound $-2h_d(x)$. Therefore, we can write 
\[
    |h_d(x)|
    \leq\frac{16 C t_d}{n}|n-2k|
    =\frac{32 C t_d}{n}\left|k-\frac n2\right|.
\]
and 
\[
\begin{aligned}
    \frac{1}{2^n}
    \sum_{x\in\mathcal B\setminus\mathcal B_{\partial}}h_d(x)^2
    &\leq
    \frac{1024 C^2 t_d^2}{n^2}
    \frac{1}{2^n}\sum_{x\in\mathcal B\setminus\mathcal B_{\partial}}
        \left(|x|-\frac n2\right)^2\\
    &\leq    
    \frac{1024 C^2 t_d^2}{n^2}
    \mathbb E_X\!\left[\left(|X|-\frac n2\right)^2\right]\\
    &=\frac{256 C^2 t_d^2}{n}.
\end{aligned}
\]
For the last equality, we have used that $|X|\sim\operatorname{Bin}(n,1/2)$ for uniformly random $X$, and hence its variance is $n/4$. Finally, we have
\[
\begin{aligned}
    \mathbb E[V_d^2]
    &=
    \frac1{2^n}\sum_{x\in\mathcal B_{\partial}}h_d(x)^2
    +
    \frac1{2^n}
    \sum_{x\in\mathcal B\setminus\mathcal B_{\partial}}h_d(x)^2\\
    &\leq
    \frac{4\Lambda_d^2t_d}{\sqrt n}
    +\frac{256 C^2 t_d^2}{n}\\
    &=O(t_d/\sqrt n),
\end{aligned}
\]
since $t_d \leq \sqrt{n}$. Together, with $\mathbb E[U_d^2]=O(t_d/\sqrt n)$, we have
that $\operatorname{Var}(\widehat p_d)=O(t_d/(m\sqrt n))$. 

\end{proof}

\section{Supporting proofs for the lower bounds}
\label{app:lower-bound-proofs}

\subsection{Proof of the selector-block lemma}
\label{app:selector-mass}
\begin{proof}[Proof of \Cref{lem:selector-blocks}]
Let $R:=2^r$ and let $T_1,\ldots,T_R$ be independently and uniformly randomly chosen $r$-element subsets of $[d]$. Each $T_j \subseteq [d]$ has exactly $r$ elements and is chosen uniformly from $\binom{d}{r}$ possible choices. We say that $y\in\{0,1\}^d$ satisfies $T_j$ if
\[
y_i=1
\qquad\text{for every }i\in T_j.
\] 
Equivalently,
\[
T_j\subseteq\operatorname{supp}(y).
\]
We define
\[
U_j
=
\left\{
y\in\{0,1\}^d:
T_j\subseteq\operatorname{supp}(y)
\text{ and }
T_k\not\subseteq\operatorname{supp}(y)
\text{ for all }k\neq j
\right\}.
\]
The sets $U_j$ are disjoint and mutually incomparable. In particular, if $y\in U_j$, $y'\in U_k$, and $j\neq k$, then we cannot have $y\leq y'$ or $y'\leq y$. If $y\leq y'$, then $T_j$ is satisfied by both $y$ and $y'$, contradicting $y'\in U_k$; the case $y'\leq y$ is symmetric.

Each $U_j$ is contained in the set of strings satisfying $T_j$. Since satisfying $T_j$ fixes the $r$ coordinates in $T_j$ to $1$, we have $|U_j| \leq 2^{d-r}$ and hence, 
\[
p_j=\frac{|U_j|}{2^d}\leq 2^{-r}.
\]
Since this holds for every $j$,
\[
p_{\max}\leq 2^{-r}.
\]

It remains to show that there exist $T_1, \ldots T_R$ such that we can lower bound the total mass
\[
\beta = \sum_{j=1}^{R} p_j = \sum_{j=1}^{R} \frac{|U_j|}{2^d}.
\]
For any fixed choice of $T_1, \ldots T_R$, $\beta$ is the probability that a uniformly random $y \in \{0,1\}^d$ belongs to exactly one of the blocks. To show that there exists a a choice of $T_1,\ldots,T_R$ for which $\beta$ is large, we will lower bound $\mathbb{E}[\beta]$ over the random choice of the sets $T_1,\ldots,T_R$.

To lower bound $\mathbb{E}[\beta]$, we identify a set $A\subseteq\{0,1\}^d$ such that $A$ contains a constant fraction of the strings $\{0, 1\}^d$. Furthermore, we will show that every $y \in A$ belongs to exactly one block with constant probability over a random choice of $T_1, \ldots T_R$. We take $A$ to be the lower half of the central layers:
\[
A
:=
\left\{
y\in\{0,1\}^d:
\frac{d}{2}-\sqrt{d}
\leq |y|
\leq \frac{d}{2}
\right\}.
\]
By Chebyshev's inequality and symmetry of the binomial distribution, a uniformly random $y\in\{0,1\}^d$ lies in $A$ with a probability at least $3/8$. It remains to bound the probability that a fixed $y\in A$ belongs to exactly one block. 

Fix such a $y$ of Hamming weight $k$. A uniformly random $r$-element set $T_j$ is contained in $\operatorname{supp}(y)$ with probability
\[
q_k
=
\frac{\binom{k}{r}}{\binom{d}{r}}.
\]
Since the $T_j$'s are chosen independently, the probability that $y$ satisfies exactly one of the sets $T_1,\ldots,T_R$, and hence belongs to exactly one block $U_j$, is
\[
R q_k(1-q_k)^{R-1}.
\]
To lower bound this quantity, we first show that $R q_k$ satisfies the bound 
\[
\frac{3}{4} \leq R q_k \leq 1
\]
for all $y \in A$. For the upper bound, note that
\[
Rq_k
=
2^r\frac{\binom{k}{r}}{\binom{d}{r}}
=
\prod_{i=0}^{r-1}\frac{2(k-i)}{d-i}.
\]
Since $k\leq d/2$, each factor in the product is at most $1$ -- therefore, we have $Rq_k\leq 1.$ For the lower bound, we may write
\[
Rq_k=\prod_{i=0}^{r-1}(1-a_i),
\qquad
a_i:=\frac{d-2k+i}{d-i}.
\]
Since $y\in A$, we have $0\leq d-2k\leq 2\sqrt{d}$.
Together with $r=\lfloor\sqrt{d}/10\rfloor$, this gives
\[
\sum_{i=0}^{r-1}a_i
\leq \frac{r(2\sqrt{d}+r)}{d-r}
\leq \frac14.
\]
Furthermore, each $a_i\in[0,1]$,
so we may use the inequality $\prod_i(1-a_i)\geq 1-\sum_i a_i$ to get
\[
Rq_k\geq 1-\sum_{i=0}^{r-1}a_i\geq \frac34.
\]
Therefore, we have shown that 
\[
\frac{3}{4} \leq R q_k \leq 1
\]
for all $y \in A$. Noting that $Rq_k \leq 1$ implies $q_k \leq \frac{1}{R}$, we can now bound 
\[
Rq_k(1-q_k)^{R-1}
\geq
\frac34
\left(1-\frac1R\right)^{R-1}
\geq
\frac{3}{4e}.
\]
It follows that, for every $y\in A$,
\[
\Pr_{T_1,\ldots,T_R}
\left[y\in\bigcup_j U_j\right]
\geq \frac{3}{4e}.
\]
Combining this with $\Pr[Y\in A]\geq3/8$, we obtain
\[
\mathbb{E}[\beta]
\geq
\Pr[Y\in A]\cdot
\inf_{y\in A}
\Pr_{T_1,\ldots,T_R}
\left[y\in\bigcup_j U_j\right]
\geq
\frac38\cdot\frac{3}{4e}
=
\frac{9}{32e}
>
\frac1{64}.
\]
Therefore, there exists a choice of $T_1,\ldots,T_R$ for which
$\beta\geq1/64$.

Finally, given a choice of $T_1, \ldots T_R$, we can define $f$ for all $y \notin \bigcup_j U_j$ in a way that preserves monotonicity.  Specifically, we let $f(y,x) = 0$ when $y$ satisfies none of the $T_j$ and $f(y,x) = 1$ when $y$ satisfies at least two of the $T_j$. As $y$ increases coordinate-wise, the set of satisfied $T_j$ can only increase. Therefore, the function remains monotone whenever each local function $f_j$ assigned to each block $U_j$ is monotone.
\end{proof}

\subsection{Proof of block factorization}
\label{app:block-isometry}
\begin{proof}[Proof of \Cref{lem:block-factorization}]
We use the local and global coherent states and the product states
defined in \Cref{subsec:block-factorization}.

We now show that there is an isometry that maps $\ket{\Phi_\lambda(f)}$ to $\ket{\chi_\lambda(f)}$
for every $f\in\mathcal F$. First, we can verify that the two families have the same pairwise inner products. For any $f,g\in\mathcal F$,
\begin{align*}
\langle\chi_\lambda(f)\mid\chi_\lambda(g)\rangle
&=
\prod_{j=1}^R
\langle
\Phi_{\lambda p_j}(f_j)
\mid
\Phi_{\lambda p_j}(g_j)
\rangle\\
&=
\prod_{j=1}^R
e^{-\lambda p_j\operatorname{dist}(f_j,g_j)}\\
&=
e^{-\lambda\operatorname{dist}(f,g)}\\
&=
\langle\Phi_\lambda(f)\mid\Phi_\lambda(g)\rangle,
\end{align*}
since we know that 
\[
\operatorname{dist}(f,g)
=
\sum_{j=1}^R p_j\operatorname{dist}(f_j,g_j).
\]
We may now construct the desired map. Define
\[
S_\lambda
:=
\operatorname{span}
\{\ket{\Phi_\lambda(f)}:f\in\mathcal F\},
\qquad
T_\lambda
:=
\operatorname{span}
\{\ket{\chi_\lambda(f)}:f\in\mathcal F\},
\]
and consider 
\[
V_\lambda
\left(
\sum_{f\in\mathcal F}
c_f\ket{\Phi_\lambda(f)}
\right)
:=
\sum_{f\in\mathcal F}
c_f\ket{\chi_\lambda(f)}.
\]
We will verify that $V_\lambda$ is well-defined,
linear, and norm preserving and hence, is an isometry.

The key observation is that the equality of the pairwise inner product tells you that
\begin{equation}
\begin{aligned}
\left\|
\sum_f c_f\ket{\Phi_\lambda(f)}
\right\|^2
&=
\sum_{f,g}\overline{c_f}c_g
\langle\Phi_\lambda(f)\mid\Phi_\lambda(g)\rangle\\
&=
\sum_{f,g}\overline{c_f}c_g
\langle\chi_\lambda(f)\mid\chi_\lambda(g)\rangle\\
&=
\left\|
\sum_f c_f\ket{\chi_\lambda(f)}
\right\|^2.
\end{aligned}
\label{eq:block-linear-combination-norms}
\end{equation}
for every choice of coefficients $(c_f)_{f\in\mathcal F}$. We now verify the three properties. 

\begin{enumerate}
\item \emph{The map is well-defined.}
Suppose that two coefficient choices represent
the same input vector:
\[
\sum_f c_f\ket{\Phi_\lambda(f)}
=
\sum_f d_f\ket{\Phi_\lambda(f)},
\]
apply \Cref{eq:block-linear-combination-norms}
to the coefficients $c_f-d_f$ gives
\[
\left\|
\sum_f(c_f-d_f)\ket{\chi_\lambda(f)}
\right\|^2
=
\left\|
\sum_f(c_f-d_f)\ket{\Phi_\lambda(f)}
\right\|^2
=
0.
\]
Hence, the output states are equal, and $V_\lambda$ is well defined.
\item \emph{Linearity.}
Linearity follows directly from the definition of the isometry. 
\item \emph{Norm preservation.}
Since every $v\in S_\lambda$ is a linear combination
of the states $\ket{\Phi_\lambda(f)}$, \cref{eq:block-linear-combination-norms}
tells us that
\[
\|V_\lambda v\|=\|v\|
\qquad\text{for every }v\in S_\lambda.
\]
\end{enumerate}

Therefore, $V_\lambda:S_\lambda\to T_\lambda$ is
an isometry satisfying \eqref{eq:coherent-block-isometry}.

We now apply the isometry to the averaged ensembles defined in
\Cref{subsec:block-factorization}.
Using \Cref{eq:coherent-block-isometry}, we have
\begin{align}
V_\lambda\Omega_b(\lambda)V_\lambda^\dagger
&=
\mathbb E_{f\sim\mathcal D_b}
\left[
\ket{\chi_\lambda(f)}\bra{\chi_\lambda(f)}
\right]
\nonumber\\
&=
\mathbb E_{(f_1,\ldots,f_R)\sim\nu_b^{\otimes R}}
\left[
\bigotimes_{j=1}^R
\ket{\Phi_{\lambda p_j}(f_j)}
\bra{\Phi_{\lambda p_j}(f_j)}
\right]
\nonumber\\
&=
\bigotimes_{j=1}^R
\mathbb E_{h\sim\nu_b}
\left[
\ket{\Phi_{\lambda p_j}(h)}
\bra{\Phi_{\lambda p_j}(h)}
\right]
\nonumber\\
&=
\bigotimes_{j=1}^R\sigma_b(\lambda p_j).
\label{eq:averaged-block-factorization}
\end{align}
for $b\in\{\textit{yes},\textit{no}\}$.

Since conjugation by an isometry preserves the
trace norm, we conclude that \eqref{eq:coherent-product-distance}
holds, completing the proof.
\end{proof}

\subsection{Proof of the local fidelity bound}
\label{app:local-fidelity-deficit}
\begin{proof}[\rev{Proof of \Cref{lem:local-fidelity-deficit}}]
\rev{We use the notation from \Cref{subsec:local-fidelity-estimate}.}

Next, recall that $\nu_b$ is the distribution over local functions as specified in in \Cref{tab:local-distributions}, for $b\in\{\textit{yes},\textit{no}\}$. 
Define $X$ and $Y$ by
\[
    X:=
    \left[
        \sqrt{\nu_{\textit{yes}}(g)}\ket{\Phi_\mu(g)}
    \right]_g,
    \qquad
    Y:=
    \left[
        \sqrt{\nu_{\textit{no}}(h)}\ket{\Phi_\mu(h)}
    \right]_h.
\]
Then,
\[
    XX^\dagger=\sigma_{\textit{yes}}(\mu),
    \qquad
    YY^\dagger=\sigma_{\textit{no}}(\mu).
\]
We have
\begin{equation}
\label{eq:mixture-fidelity-overlap}
    F_\mu =\|X^\dagger Y\|_1,
\end{equation}
where the entries of $X^\dagger Y$ are
\[
    (X^\dagger Y)_{g,h}
    =
    \sqrt{\nu_{\textit{yes}}(g)\nu_{\textit{no}}(h)}
    \,\langle\Phi_\mu(g)\mid\Phi_\mu(h)\rangle.
\]

Set $\widetilde{\mu}:=\frac{\mu p}{2}$ and define
\[
    \ket{\Gamma_\mu(g)}
    :=
    e^{-\widetilde{\mu}/2}
    \bigoplus_{k=0}^{\infty}
    \frac{\widetilde{\mu}^{k/2}}{\sqrt{k!}}
    \ket{z_g}^{\otimes k},
\]
where the $k=0$ sector is some fixed unit vector $\ket{\emptyset}$, independent of $g$. We have that 
\[
\begin{aligned}
    \langle\Gamma_\mu(g)\mid\Gamma_\mu(h)\rangle
    &=
    e^{-\widetilde{\mu}}
    \sum_{k=0}^{\infty}
    \frac{\widetilde{\mu}^k}{k!}
    {\langle z_g\mid z_h\rangle}^k\\
    &=
    \exp\!\left[
        -\widetilde{\mu}
        \bigl(1-\langle z_g\mid z_h\rangle\bigr)
    \right]\\
    &=
    e^{-\mu\operatorname{dist}(g,h)}\\
    &=
    \langle\Phi_\mu(g)\mid\Phi_\mu(h)\rangle,
\end{aligned}
\]
where the last equality follows from \Cref{eq:overlap_between_phi}.
Therefore, every entry of $X^\dagger Y$ can be written as
\[
    (X^\dagger Y)_{g,h}
    =
    \sqrt{\nu_{\textit{yes}}(g)\nu_{\textit{no}}(h)}
    \,\langle\Gamma_\mu(g)\mid\Gamma_\mu(h)\rangle.
\]
In particular, replacing the component vectors
$\ket{\Phi_\mu(g)}$ by $\ket{\Gamma_\mu(g)}$ leaves the overlap matrix and hence, the trace norm $F_{\mu}$ unchanged. 

Next, we will retain only the $k=0$ and $k=1$ terms of $\Gamma_{\mu}(g)$. Define
\[
    \ket{\widetilde\Gamma_\mu(g)}
    :=
    \frac{
        \ket{\emptyset}
        +\sqrt{\widetilde{\mu}}\ket{z_g}
    }{\sqrt{1+\widetilde{\mu}}},
\]
and define the corresponding mixed states as
\begin{equation}
\label{eq:definition_of_tau_beta_mu}
    \tau_b(\mu)
    :=
    \sum_g \nu_b(g)
    \ket{\widetilde\Gamma_\mu(g)}
    \bra{\widetilde\Gamma_\mu(g)},
    \qquad
    b\in\{\textit{yes},\textit{no}\}.
\end{equation}
The total squared norm of the discarded terms is
independent of $g$ and satisfies
\begin{equation}
\label{eq:coherent-truncation-tail}
\begin{aligned}
    q_{\mathrm{tail}}
    &:=
    e^{-\widetilde{\mu}}
    \sum_{k=2}^{\infty}
    \frac{\widetilde{\mu}^k}{k!}\\
    &\leq
    \frac{\widetilde{\mu}^2}{2}\,
    e^{-\widetilde{\mu}}
    \sum_{k=2}^{\infty}
    \frac{\widetilde{\mu}^{k-2}}{(k-2)!}\\
    &=
    \frac{\widetilde{\mu}^2}{2}.
\end{aligned}
\end{equation}
We bound the effect of this truncation on the fidelity. If $\widetilde{\mu} = 0$, then the truncation has no effect. For $\widetilde{\mu}>0$, we may write
\[
    \ket{\Gamma_\mu(g)}
    =
    \sqrt{1-q_{\mathrm{tail}}}\,
    \ket{\widetilde\Gamma_\mu(g)}
    +
    \sqrt{q_{\mathrm{tail}}}\,\ket{\eta_g},
\]
where $\ket{\eta_g}$ is the normalized discarded terms supported on $k \geq 2$. Using this decomposition, we obtain
\[
\begin{aligned}
    (X^\dagger Y)_{g,h}
    ={}&
    (1-q_{\mathrm{tail}})
    \sqrt{\nu_{\textit{yes}}(g)\nu_{\textit{no}}(h)}
    \,\langle\widetilde\Gamma_\mu(g)\mid
    \widetilde\Gamma_\mu(h)\rangle  +
    q_{\mathrm{tail}}
    \sqrt{\nu_{\textit{yes}}(g)\nu_{\textit{no}}(h)}
    \,\langle\eta_g\mid\eta_h\rangle,
\end{aligned}
\]
since $\langle\widetilde\Gamma_\mu(g)\mid\eta_h\rangle=0$
for all $g,h$.

We can think of $X^\dagger Y$ as the sum of two matrices: 
\[
\begin{aligned}
    (1-q_{\mathrm{tail}})
    \left[
        \sqrt{\nu_{\textit{yes}}(g)\nu_{\textit{no}}(h)}
        \,\langle\widetilde\Gamma_\mu(g)\mid
                 \widetilde\Gamma_\mu(h)\rangle
    \right]_{g,h} + 
    q_{\mathrm{tail}}
    \left[
        \sqrt{\nu_{\textit{yes}}(g)\nu_{\textit{no}}(h)}
        \,\langle\eta_g\mid\eta_h\rangle
    \right]_{g,h}.
\end{aligned}
\]
The first matrix is the weighted overlaps
of the vectors $\tau_{\textit{yes}}(\mu)$ and
$\tau_{\textit{no}}(\mu)$ defined in
\Cref{eq:definition_of_tau_beta_mu}. Its trace norm equals $F(\tau_{\textit{yes}}(\mu),\tau_{\textit{no}}(\mu))$. The trace norm of the second matrix is equal to the fidelity between 
\[
    \sum_g\nu_{\textit{yes}}(g)\ket{\eta_g}\bra{\eta_g},
    \qquad
    \sum_h\nu_{\textit{no}}(h)\ket{\eta_h}\bra{\eta_h},
\]
and can be bound by $1$. Applying the reverse triangle inequality, we get
\begin{equation}
\label{eq:local-fidelity-truncation}
\begin{aligned}
    F_\mu
    &\geq
    (1-q_{\mathrm{tail}})
    F(\tau_{\textit{yes}}(\mu),\tau_{\textit{no}}(\mu))
    -q_{\mathrm{tail}}\\
    &\geq
    F(\tau_{\textit{yes}}(\mu),\tau_{\textit{no}}(\mu))
    -2q_{\mathrm{tail}}\\
    &\geq
    F(\tau_{\textit{yes}}(\mu),\tau_{\textit{no}}(\mu))
    -\frac{\mu^2p^2}{4},
\end{aligned}
\end{equation}
where we used that $F(\tau_{\textit{yes}}(\mu),\tau_{\textit{no}}(\mu))$ is bound by $1$ and \Cref{eq:coherent-truncation-tail}. 

We now bound $F(\tau_{\textit{yes}}(\mu),\tau_{\textit{no}}(\mu))$. Inside $B$, the local functions satisfy 
$f_i^-=1-f_i^+$ and $h_1=1-h_0$ so we may write
\[
    \ket{z_{f_i^-}}=-\ket{z_{f_i^+}},
    \qquad
    \ket{z_{h_1}}=-\ket{z_{h_0}}.
\]
Using \Cref{eq:definition_of_tau_beta_mu}, we have
\[
\begin{aligned}
    \tau_{\textit{yes}}(\mu)
    &=
    \frac{\alpha}{m}\sum_{i=1}^m
    \ket{\widetilde\Gamma_\mu(f_i^+)}
    \bra{\widetilde\Gamma_\mu(f_i^+)}
    +
    \frac{1-\alpha}{2}\sum_{s=0}^1
    \ket{\widetilde\Gamma_\mu(h_s)}
    \bra{\widetilde\Gamma_\mu(h_s)}\\
    &=
    \frac{1}{1+\widetilde{\mu}}
    \Biggl[
    \frac{\alpha}{m}\sum_{i=1}^m
    \bigl(
        \ket{\emptyset}
        +\sqrt{\widetilde{\mu}}\ket{z_{f_i^+}}
    \bigr)
    \bigl(
        \bra{\emptyset}
        +\sqrt{\widetilde{\mu}}\bra{z_{f_i^+}}
    \bigr)\\
    &\qquad\qquad+
    \frac{1-\alpha}{2}\sum_{s=0}^1
    \bigl(
        \ket{\emptyset}
        +\sqrt{\widetilde{\mu}}\ket{z_{h_s}}
    \bigr)
    \bigl(
        \bra{\emptyset}
        +\sqrt{\widetilde{\mu}}\bra{z_{h_s}}
    \bigr)
    \Biggr]\\
    &=
    \frac{1}{1+\widetilde{\mu}}
    \Bigl[
        \ket{\emptyset}\bra{\emptyset}
        +\widetilde{\mu}C
        +\alpha\sqrt{\widetilde{\mu}}
        \bigl( \ket{\emptyset}\bra{\bar z}
            +\ket{\bar z}\bra{\emptyset}
        \bigr)
    \Bigr].
\end{aligned}
\]
The last equality uses $\ket{z_{h_1}}=-\ket{z_{h_0}}$ and the definitions of $\ket{\bar z}$ and $C$. Similarly,
\[
    \tau_{\textit{no}}(\mu)
    =
    \frac{1}{1+\widetilde{\mu}}
    \Bigl[
        \ket{\emptyset}\bra{\emptyset}
        +\widetilde{\mu}C
        +(\alpha-2\rho)\sqrt{\widetilde{\mu}}
        \bigl(
            \ket{\emptyset}\bra{\bar z}
            +\ket{\bar z}\bra{\emptyset}
        \bigr)
    \Bigr].
\]
Notice that only the cross terms between $\ket{\emptyset}$ and $\ket{\bar z}$ differ. Define
\[
    R:=
    \widetilde{\mu}
    \left(C-\alpha\ket{\bar z}\bra{\bar z}\right),
    \qquad
    u:=\widetilde{\mu}\|\bar z\|^2.
\]
Since $R\ket{\emptyset}=R\ket{\bar z}=0$ and $R$ is
Hermitian, $R$ is supported on the orthogonal complement
of their span. The remaining terms act in the two-dimensional space spanned by $\ket{\emptyset}$ and $\ket{\bar z}$. In the basis $\{\ket{\emptyset},\ket{\bar z}/\|\bar z\|\}$, they give the blocks
\[
    A_{\textit{yes}}
    :=
    \begin{pmatrix}
        1 & \alpha\sqrt u\\
        \alpha\sqrt u & \alpha u
    \end{pmatrix},
    \qquad
    A_{\textit{no}}
    :=
    \begin{pmatrix}
        1 & (\alpha-2\rho)\sqrt u\\
        (\alpha-2\rho)\sqrt u & \alpha u
    \end{pmatrix}.
\]
Therefore, we may write
\[
    \tau_{\textit{yes}}(\mu) = \frac{A_{\textit{yes}}\oplus R}{1+\widetilde{\mu}},
    \qquad
    \tau_{\textit{no}}(\mu) = \frac{A_{\textit{no}}\oplus R}{1+\widetilde{\mu}}.
\]

Since fidelity adds across orthogonal blocks and $F(R,R)=\operatorname{Tr}R=\widetilde{\mu}-\alpha u$, we have
\begin{equation}
\label{eq:truncated-fidelity-blocks}
    F(\tau_{\textit{yes}}(\mu),\tau_{\textit{no}}(\mu))
    =
    \frac{
        \widetilde{\mu}-\alpha u
        +F(A_{\textit{yes}},A_{\textit{no}})
    }{1+\widetilde{\mu}}.
\end{equation} For positive semidefinite $2\times2$ matrices, root fidelity satisfies
\[
    F(A,B)^2
    =
    \operatorname{Tr}(AB)+2\sqrt{\det A\,\det B}.
\]
For our blocks,
\begin{align*}
    \operatorname{Tr}(A_{\textit{yes}}A_{\textit{no}})
    &=
    1+2\alpha(\alpha-2\rho)u+\alpha^2u^2,\\
    \det A_{\textit{yes}}
    &=
    \alpha(1-\alpha)u,\\
    \det A_{\textit{no}}
    &=
    \bigl[\alpha(1-\alpha)+4\rho(\alpha-\rho)\bigr]u.
\end{align*}
We obtain 
\begin{equation}
\label{eq:explicit-block-fidelity}
\begin{aligned}
    F(A_{\textit{yes}},A_{\textit{no}})
    =\Bigl[
        &1+2\alpha(\alpha-2\rho)u+\alpha^2u^2 +2u\sqrt{
            \alpha(1-\alpha)
            \bigl[\alpha(1-\alpha)+4\rho(\alpha-\rho)\bigr]
        }
    \Bigr]^{1/2}.
\end{aligned}
\end{equation}
It remains to bound $1-F_\mu$ for all $j$. If $\rho = 0$, the local distributions coincide and $F_{\mu} = 1$. Assume $0 <\rho\leq\alpha\leq1$, and let $F_A:=F(A_{\textit{yes}},A_{\textit{no}})$. By \cref{eq:local-fidelity-truncation} and \cref{eq:truncated-fidelity-blocks}, we get 
\begin{align*}
1-F_\mu
&\leq
1-F(\tau_{\textit{yes}}(\mu),\tau_{\textit{no}}(\mu))
+\frac{\mu^2p^2}{4}\\
&=
\frac{1+\alpha u-F_A}{1+\widetilde{\mu}}
+\frac{\mu^2p^2}{4}\\
&=
\frac{(1+\alpha u)^2-F_A^2}
     {(1+\widetilde{\mu})(1+\alpha u+F_A)}
+\frac{\mu^2p^2}{4}.
\end{align*}
We now upper bound $(1+\alpha u)^2-F_A^2$ and lower bound 
$1+\alpha u+F_A$. \cref{eq:explicit-block-fidelity} gives
\begin{align*}
(1+\alpha u)^2-F_A^2
&=
2u\left[
\alpha(1-\alpha)+2\alpha\rho
-\sqrt{
\alpha(1-\alpha)
\bigl[\alpha(1-\alpha)+4\rho(\alpha-\rho)\bigr]
}
\right]\\
&=
\frac{8\alpha\rho^2u}{
\alpha(1-\alpha)+2\alpha\rho
+\sqrt{
\alpha(1-\alpha)
\bigl[\alpha(1-\alpha)+4\rho(\alpha-\rho)\bigr]
}
}\\
&\leq
\frac{8\alpha\rho^2u}
     {2\alpha(1-\alpha)+2\alpha\rho}\\
&=
\frac{4\rho^2u}{1-\alpha+\rho},
\end{align*}
For the denominator, \cref{eq:explicit-block-fidelity} gives 
\begin{align*}
F_A^2
&\geq
1+2\alpha(\alpha-2\rho)u+\alpha^2u^2\\
&=
(1-\alpha u)^2+2\alpha(1+\alpha-2\rho)u\\
&\geq
(1-\alpha u)^2.
\end{align*}
Hence, we have $F_A \geq |1-\alpha u|$ which means $1+\alpha u+F_A \geq 2$. Combining these bounds, we have
\begin{equation}
\label{eq:local-fidelity-deficit}
\begin{aligned}
1-F_\mu
&\leq
\frac{2\rho^2u}
     {(1-\alpha+\rho)(1+\widetilde{\mu})}
+\frac{\mu^2p^2}{4}\\
&\leq
\frac{\rho^2}{1-\alpha+\rho}\,
\mu p\|\bar z\|^2
+\frac{\mu^2p^2}{4},
\end{aligned}
\end{equation}
where we have substituted
$u=\mu p\|\bar z\|^2/2$ and
$1+\widetilde{\mu}\geq1$. \end{proof}

\end{document}